# 3D Topologically Polarized Elastic Metamaterials Enable Asymmetric Energy Isolation at Low Frequencies

Shaoyuan Zhang[1]†, Xuejian Gong[1]†, Fangyuan Ma[1], Zheng Tang[1], Ying Wu[2]*, Di Zhou[1]*, Feng Li[1]*, and Yugui Yao[1]

[1]Key Lab of Advanced Optoelectronic Quantum Architecture and Measurement (MOE), School of Physics, Beijing Institute of Technology, Beijing 100081, China

[2]School of Physics, Nanjing University of Science and Technology, Nanjing 210094, China

**Topologically polarized elasticity has been extensively studied in lower-dimensions, yet its three-dimensional (3D) counterpart remains largely unexplored. Here, we demonstrate omnidirectional topological elasticity in 3D structures that incorporate bending stiffness, which elevates zero-frequency topological mechanical states into finite-frequency phononic modes. These modes are localized at a single boundary, creating a pronounced stiffness contrast in both static and finite-frequency dynamic regimes. This three-dimensional structure exhibits highly polarized mechanical behavior across all spatial dimensions, establishing omnidirectional asymmetric topological elasticity. Experimental and numerical results confirm robust, asymmetric energy isolation, arising from the interplay between bulk topological polarization and boundary-localized surface modes. Our findings establish a paradigm for 3D metamaterials, with promising applications in vibration shielding and directional wave manipulation.**

†These authors contributed equally to this work.

*Correspondence author: yingwu@njust.edu.cn, dizhou@bit.edu.cn, phlifeng@bit.edu.cn

## Introduction

The application of quantum topological band theory (*1-3*) to classical structures (*4-9*) has given rise to the field of topological metamaterials (*10-13*), in which nontrivial band topologies of classical fields enable unique wave dynamics and control (*14-17*). Using current 3D-printing technology, one can design artificial atoms to realize a broad range of emerging applications, including topological chiral edge waves (*18*), unidirectional light transport (*19*), and topological heat transfer (*20*). Among the diverse classes of topological mechanical systems, a particularly intriguing category is that of isostatic structures, whose bulk features a precise balance between degrees of freedom and constraints, placing them at the brink of mechanical instability *(21-26)*. Such isostatic networks host zero-frequency topological modes that dictate the mechanical softness at their open boundaries.

Recent advances have enabled the manipulation of finite-frequency topological phononic modes in one- and two-dimensional (2D) continuum metamaterials, leading to asymmetric boundary elasticity (*27-30*). Unlike discrete isostatic frames with friction-based hinges, these continuum metamaterials incorporate finite thickness ligaments that inherently support bending. This bending stiffness minimizes energy loss from frictional dissipation and shifts the characteristic zero frequency topological states of isostatic systems to small but finite frequencies. As a result, these materials gain enhanced mechanical stability while enabling low-loss waveguiding. Despite this frequency shift, the soft modes retain their topological protection (*31,32*), enabling asymmetric sound isolation. Moreover, omnidirectional control of phononic modes has been demonstrated in bi-layered two-dimensional topological lattices (*33*). However, 1D and 2D architectures face a fundamental challenge: their susceptibility to buckling under out-of-plane deformations. This structural instability undermines the reliability of topological mechanical protection and, consequently, imposes limitations on practical applications.

Here, we address these challenges by implementing topologically polarized elasticity in three-dimensional elastic metamaterials, where soft modes localize exclusively at one surface and the opposing surface remains rigid, establishing a distinct elastic contrast. Incorporating small but finite bending stiffness achieves two key objectives: (1) elevating zero-frequency topological edge states into finite-frequency boundary modes, and (2) ensuring that such topological elastic modes emerge exclusively at the soft boundary, with none appearing at the rigid boundary.

Furthermore, we demonstrate strongly asymmetric wave propagation governed by boundary type. Excitation at soft boundaries confines energy to the surface, whereas input at rigid boundaries allows waves to penetrate the bulk and transmit across the boundary. This directional contrast is most pronounced when the excitation frequency overlaps with the topological soft-mode spectrum. By incorporating geometric topology into 3D material design, our framework enables systematic control of quasi-static responses and offers a versatile route to asymmetric wave isolation.

## Results

The edge vectors of the 12 bonds are determined by connecting the nearest-neighbor nodes, as detailed in Fig. S1 of the Supplementary Material. For the spring-mass model in Fig. 1(**A**), each node represents a mass particle and each edge acts as a Hookean spring. In the pyrochlore structure, 12 constraints are imposed on the unit cell, exactly balancing its $4 \times 3 = 12$ degrees of freedom and yielding an isostatic lattice (*34*). As a result, boundary conditions play a crucial role in shaping its mechanical behavior. Switching from periodic to open boundaries removes three constraints per supercell, leading to the emergence of three zero-frequency mechanical states.

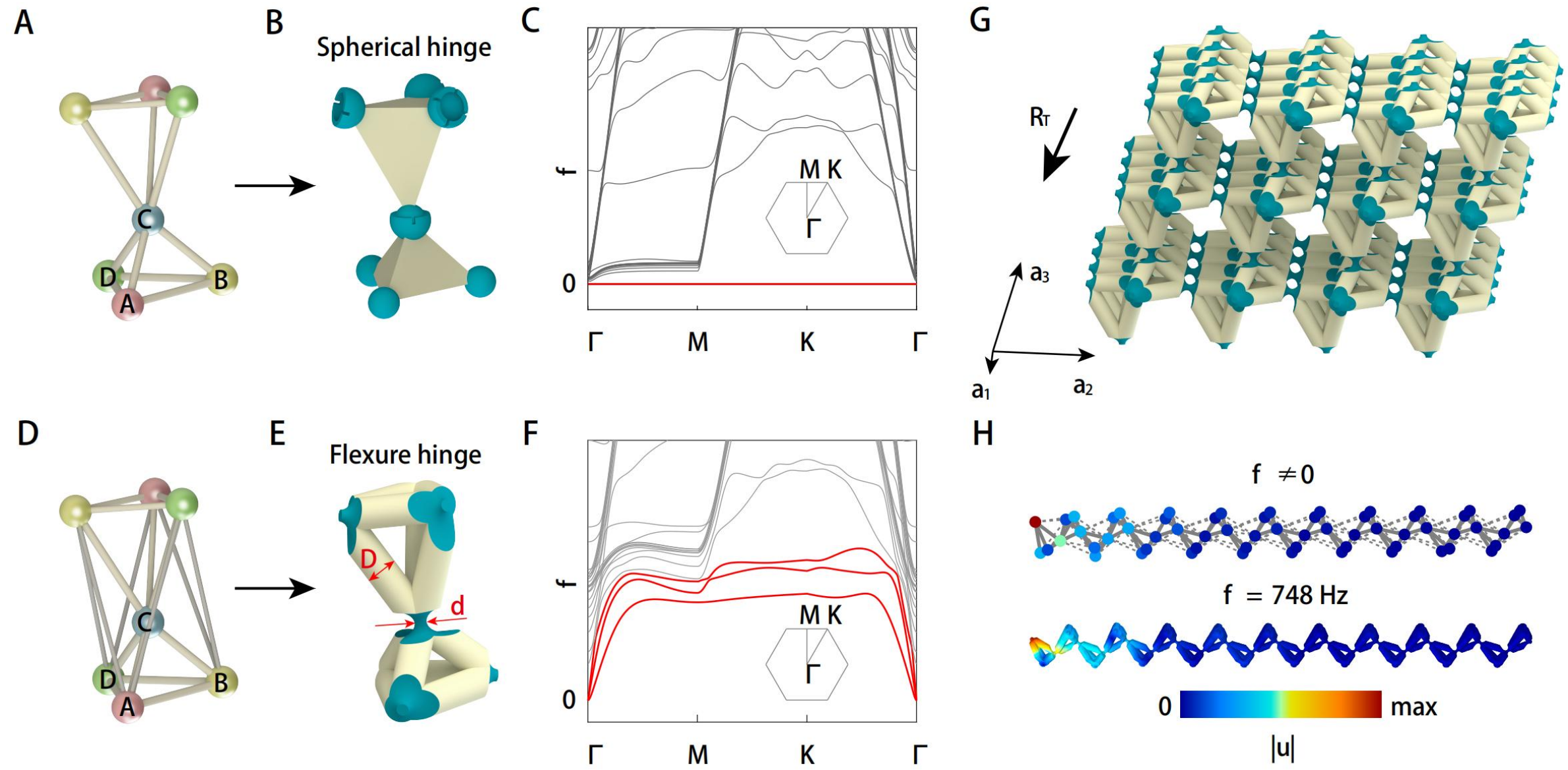

**FIG. 1. Geometric Design and Surface Modes of Topologically Polarized Elastic Metamaterials.** (**A**) Unit cell of the generalized pyrochlore lattice in a spring-mass model at the isostatic point (m = 2.3 g, $k = 1 \times 10^5$ N/m). The unit cell consists of four sites and 12 edges, with the node positions $\boldsymbol{A}$ = (0.5530, 0.5159, 0) $\ell$, $\boldsymbol{B}$ = (0.0029, 0.5000, 0.5530) $\ell$, $\boldsymbol{C}$ = (0.4470, 0.1000, 0.4470) $\ell$, and $\boldsymbol{D}$ = (−0.0954, −0.1000, 0.0636) $\ell$, where $\ell$ = 24 *mm*. The primitive vectors of the deformed pyrochlore lattice are $\boldsymbol{a}_1$ = (1, 1, 0) $\ell$, $\boldsymbol{a}_2$ = (0, 1, 1) $\ell$, and $\boldsymbol{a}_3$ = (1, 0, 1) $\ell$. (**B**) Schematic of rigid tetrahedral units connected by spherical hinges, forming an isostatic lattice. (**C**) Phononic band structure of the model in **(A)**, showing three zero-frequency topological surface modes (red flat curves). (**D**) Modified spring-mass model with next-nearest-neighbor (NNN) bonds ($\kappa = 6.5 \times 10^2$ N/m) to mimic bending stiffness. (**E**) Blueprint of the 3D-printed unit cell with edges of diameter D = 4.8 mm and hinges of diameter d = 1.5 mm, emulating NNN softness. (**F**) Dispersion relation of the model in **(D)**, showing three topological soft modes (red curves). (**G**) Schematic of the printed lattice with soft hinges; polarization vector $\boldsymbol{R}_T$ points downward, yielding soft bottom surface. Inset: primitive lattice vectors. (**H**) Top: one topological surface mode from (**D**). Bottom: Finite element simulation of the topological mode in **(E)** at 748 Hz.

The static mechanics of this spring-mass model is governed by the compatibility matrix $\mathbf{C}$, which relates node displacements to edge elongations. In crystalline structures, $\mathbf{C}$ can be expressed in reciprocal space as $\mathbf{C}(\boldsymbol{k})$, with $\boldsymbol{k}$ denoting the wavevector in the Brillouin zone. From $\mathbf{C}(\boldsymbol{k})$, one defines integer-valued winding numbers $N_i(\boldsymbol{k}) =$ $\frac{-1}{2\pi}\oint_{\boldsymbol{k}\to\boldsymbol{k}+\boldsymbol{b}_i} d\boldsymbol{k}\cdot\nabla_k \mathrm{Im}\ln\det \mathbf{C}(\boldsymbol{k})$ that characterize the topological phase of the mechanical system. These invariants are obtained by integrating along closed loops $\boldsymbol{k}\to\boldsymbol{k}+\boldsymbol{b}_i$ parallel to the reciprocal vectors $\boldsymbol{b}_i$, subject to the relation $\boldsymbol{a}_i\cdot\boldsymbol{b}_j = 2\pi\delta_{ij}$ , where $\boldsymbol{a}_i$ ( $i = 1,2,3$ ) denote the primitive vectors. In general, winding numbers can manifest qualitatively distinct behaviors depending on the unit cell geometry. For most geometric configurations of pyrochlore structures, the winding numbers exhibit integer jumps as the wavevector traverses the Brillouin zone, with the critical boundaries between distinct integer regions identified as mechanical Weyl lines (*35*).

However, in models with gapped mechanical spectra, such as the structure shown in Fig.1, the winding numbers remain invariant for arbitrary wavevectors, and the mechanical bands contain no Weyl lines. These globally defined invariants collectively form the topological polarization vector (*34, 36, 37*): $\boldsymbol{R}_T = \sum_{i=1}^{3} N_i\,\boldsymbol{a}_i$, which serves as a characteristic descriptor of the topological phase for zero-frequency mechanical modes. These modes exhibit selective localization at the open boundary where the topological polarization terminates, resulting in pronounced static softness. In contrast, the opposing boundary remains rigid due to the absence of such topological polarization. In principle, this mechanically isostatic structure and the corresponding zero-frequency topological states can be represented schematically by incorporating spherical hinges (*34*) within rigid tetrahedral units, as illustrated in Fig. 1(**B**). While the dynamic behaviors differ between the theoretical spring-mass model and the 3D-printed spherical-hinge architecture, their static mechanical responses remain consistent. In particular, both exhibit topological static states localized at a single boundary, because the static mechanical modes preserve the rigid bodies and edge lengths across both models. It is worth noting that although specific parameter values are used in Fig. 1, the polarized elasticity is protected by the system's topological property in its band structure. As long as the phononic bandgap remains open (i.e., the mechanical spectrum contains no Weyl lines), this robust and polarized elasticity persists and does not depend on the specific geometric parameters chosen.

In the current geometric configuration, shown in Figs. 1(**A**, **B**), the topological polarization is given by $\boldsymbol{R}_T = \boldsymbol{a}_1 - \boldsymbol{a}_2 - 2\boldsymbol{a}_3$ , indicating that the three zero-energy mechanical modes per supercell are topologically polarized and localize on the bottom boundary of this isostatic structure (see Supplementary Material). The zero-frequency nature of the topological boundary states leads to a vanishing group velocity [see the red flat line in Fig. 1(**C**)]. As a result, these modes remain stationary and cannot support wave propagation.

Based on the static topological mechanics in the isostatic spring-mass model, we examine how surface modes evolve as additional interactions disrupt isostaticity. Theoretically, we introduce next-nearest-neighbor (NNN) bonds into the deformed pyrochlore lattice, as shown in Fig. 1(**D**), with a spring stiffness of $\kappa = 6.5 \times 10^{-3}k$, where $k$ is the stiffness of the NN bonds. All parameters in the spring-mass model are chosen to fit the experimental data. These NNN bonds emulate bending stiffness, thereby offering insight into super-isostatic elastic behavior (*27, 30*). This equivalence arises because, within linear elasticity, bending can be expressed as the differential displacement across three adjacent points — effectively introducing next-nearest-neighbor interactions (see Fig. S2, Supplementary Material). These NNN bonds slightly modify the phononic bulk band structure while substantially elevating the topological surface states to finite frequencies. The frequency increase of topological modes is intuitive, as they originate from zero-frequency states in the absence of NNN interactions. As the NNN stiffness $\kappa$ increases, the eigenfrequencies and group velocities of topological surface modes scale proportionally to $\kappa^{\frac{1}{2}}$. In Fig. 1(**E**), this design principle is illustrated schematically by small, finite hinges connecting rigid tetrahedral units, forming a blueprint that intrinsically introduces bending stiffness into the structure.

As depicted in Figs. 1(**F**), the band structure for the topological soft modes features three lowest-frequency branches, each confined to intervals below the dispersive region of the lowest bulk modes. These three topological soft modes arise from network connectivity, as three constraints are released during the transition from periodic to open boundary conditions. Their spectral separation from bulk acoustic dispersions preserves topological robustness (*31*) and maintains surface localization, even in the presence of structural disorder. This contrasts sharply with conventional Rayleigh waves, which typically support a single boundary mode, demonstrating the fundamental difference between topologically polarized soft modes and classical surface waves.

Building upon the established theoretical framework, we now investigate and implement topologically polarized elasticity in our 3D-printed metamaterial. As shown in Fig. 1(**G**), the finite structure adopts the geometric configuration of a topologically polarized pyrochlore lattice. Each unit cell consists of thin, finite ligaments linking two tetrahedra. Specifically, the corner positions of the left tetrahedron are positioned at $\boldsymbol{A}$, $\boldsymbol{B}$, $\boldsymbol{C}$, and $\boldsymbol{D}$, whereas the nodes of the right tetrahedron are located at $\boldsymbol{A}' = \boldsymbol{A} - \boldsymbol{a}_1 + \boldsymbol{a}_3$, $\boldsymbol{B}' = \boldsymbol{B} - \boldsymbol{a}_2 + \boldsymbol{a}_3$, $\boldsymbol{C}' = \boldsymbol{C}$, and $\boldsymbol{D}' = \boldsymbol{D} + \boldsymbol{a}_3$. Finally, the two tetrahedra are joined at tip $\boldsymbol{D}$ to constitute the unit cell. The geometric parameters of the bonds and ligaments are depicted in Fig. 1(**E**), with $D$ = 4.8 mm; $d$ = 1.5 mm (see Fig. S6, Supplementary Material). These ligaments introduce a small but finite bending stiffness in the soft metamaterial. The lattice comprises 15×15×10 unit cells, subjected to open boundary conditions on surfaces normal to the reciprocal vector $\boldsymbol{b}_3$, while the four side boundaries are held fixed.

The realistic elastic structure, with ligaments of small but finite thickness, exhibits a nonzero bending stiffness, shifting the metamaterial away from the isostatic point. These finite ligaments enhance mechanical stability, preventing structural failure induced by zero-frequency mechanical modes and raising their frequencies to finite values. Akin to the topologically polarized soft modes in the theoretical spring-mass model that is beyond the isostatic limit [top panel of Fig. 1(**H**)], soft modes in this 3D-printed metamaterial also exhibit topological polarization. This is demonstrated numerically via finite element analysis, which reveals that the soft modes are localized at the left soft boundary [lower panel of Fig. 1(**H**)], and are characterized by the topological polarization vector $\boldsymbol{R}_{\mathrm{T}}$. Owning to their nonzero frequency and group velocity, these numerically computed topological modes offer an effective mechanism for asymmetric isolation of elastic modes, which we leverage and experimentally implement in both the static and dynamic response of our 3D-printed metamaterial.

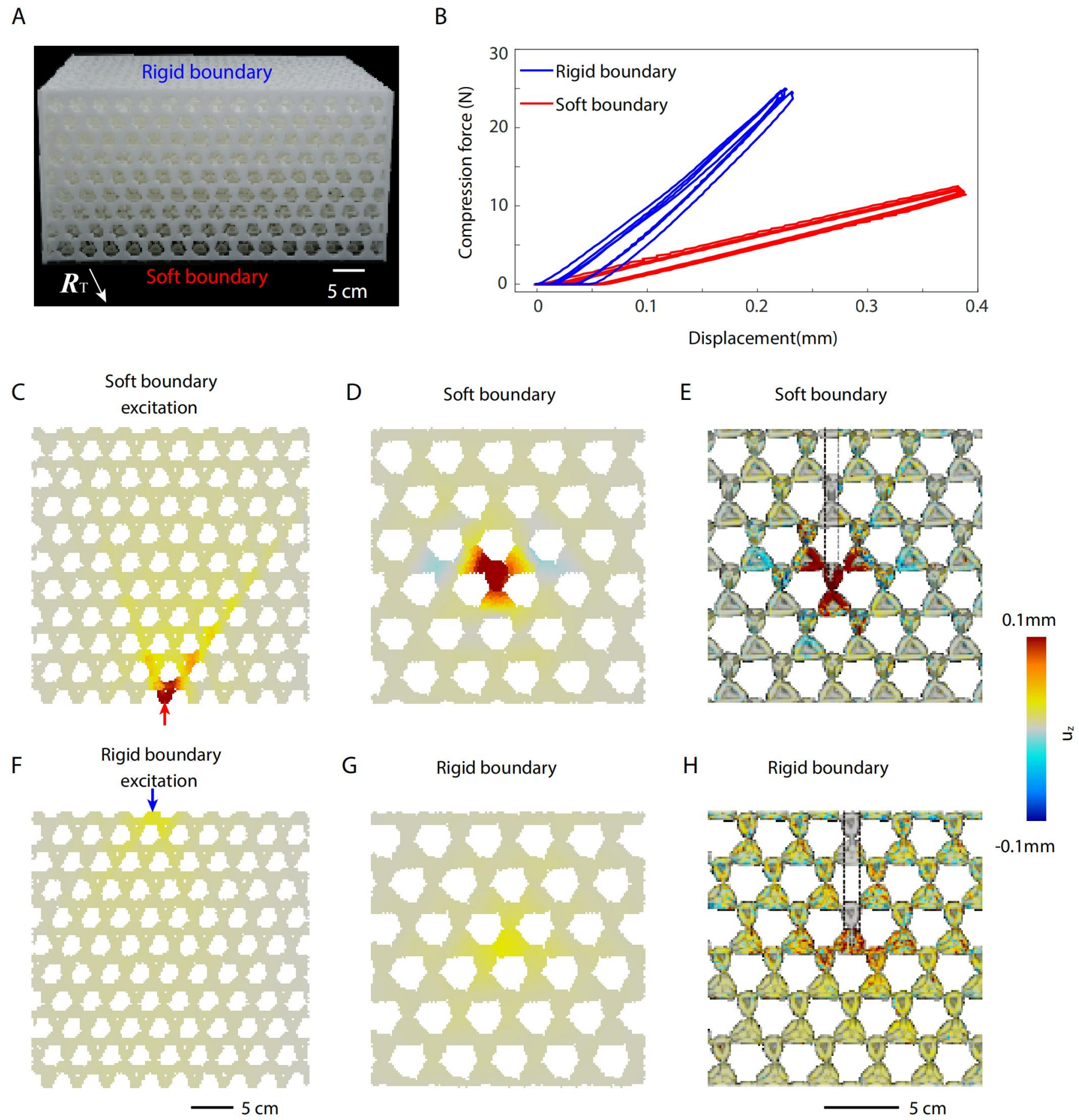

**FIG. 2. Static Topological Mechanics in the 3D-Printed Elastic Metamaterial.** (**A**) Photograph of the 3D-printed metamaterial, highlighting the topological polarization vector and indicating the soft and rigid boundaries. (**B**) Force–displacement measurements over four loading cycles used to characterize the boundary stiffness at the soft and rigid surfaces. (**C**) (**D**) and (**F**) (**G**) Numerical simulations of surface deformation under static, normal inward forces, applied at the soft boundary (**C, D**) and the rigid boundary (**F, G**), with results shown from side and top views. (**E**) (**H**) Experimental top views of surface deformation under static loading at the soft (**E**) and rigid (**H**) boundaries, further validating the predictions in (**C, D**) and (**F, G**), respectively. The dashed box represents the rod where the excitation source is placed.

As shown in Fig. 2(**A**), the sample is fabricated via 3D printing with photosensitive resin, characterized by a Young's modulus of $E = 2.6$ GPa, Poisson's ratio $\nu = 0.3$, and density $\rho = 1050\ \mathrm{kg/m^3}$ . Fig. 2(**B**) shows surface deformation responses under a quasistatically applied normal force ranging from 0 to 25 N exerted on the soft and rigid boundaries, respectively. The topologically soft surface undergoes a highly localized and concave deformation, while the topologically rigid boundary exhibits a uniform and global elastic deformation constrained by the fixed boundaries on all four sides of the metamaterial. Notably, the rigid surface demonstrates approximately four times the rigidity of the soft boundary, quantitatively confirming the presence of highly polarized static boundary elasticity. This stiffness contrast diminishes as the bending stiffness increases. The observed hysteresis originates from the viscoelastic nature of the plastic material.

Figs. 2(**C**) and 2(**D**) display the side and top views of elastic deformation resulting from a static normal force of 10 N applied to the soft boundary, with the experimental setup detailed in Fig. S7 of the Supplementary Material. Experimental data obtained via laser scanning in Fig. 2(**E**) confirm that the soft boundary exhibits localized deformation under loading. In contrast, Figs. 2(**F**) and 2(**G**) present numerical simulations of lattice deformation under the same inward static normal force applied to the rigid boundary. Corresponding experimental measurements in Fig. 2(**H**) further verify that the rigid boundary undergoes uniform deformation. The static nature of this elastic asymmetry originates from the global topological geometry of the metamaterial, distinguishing it from existing works in which conduction bands persist below the lower bound of the working frequency range for mode isolation.

To further investigate the dynamical responses of the elastic metamaterial, we apply an external vibrational excitation at the central point of either the soft or rigid surface (see Fig. S8 in Supplementary Materials for experimental setup) and perform the experimental measurement shown in Fig. S9. Figs. 3(**A**, **B**) present the experimental band structures of the 3D-printed metamaterial, with black lines denoting numerical results that exhibit excellent agreement. These results further reveal distinct asymmetric elastic dynamics between Figs. 3(**A**) and (**B**). In Fig. 3(**A**), the soft boundary supports three lowest frequency dispersion curves, indicating the presence of topological soft acoustic modes, a feature that is absent in Fig. 3(**B**) for the rigid boundary. The dispersion relation further demonstrates that these surface modes acquire a finite group velocity, enabling propagation along the lattice boundary. The analytical results for the dispersion curves corresponding to Figs. 3(**A**) and 3(**B**) are shown in Fig. S3 of the Supplementary Material.

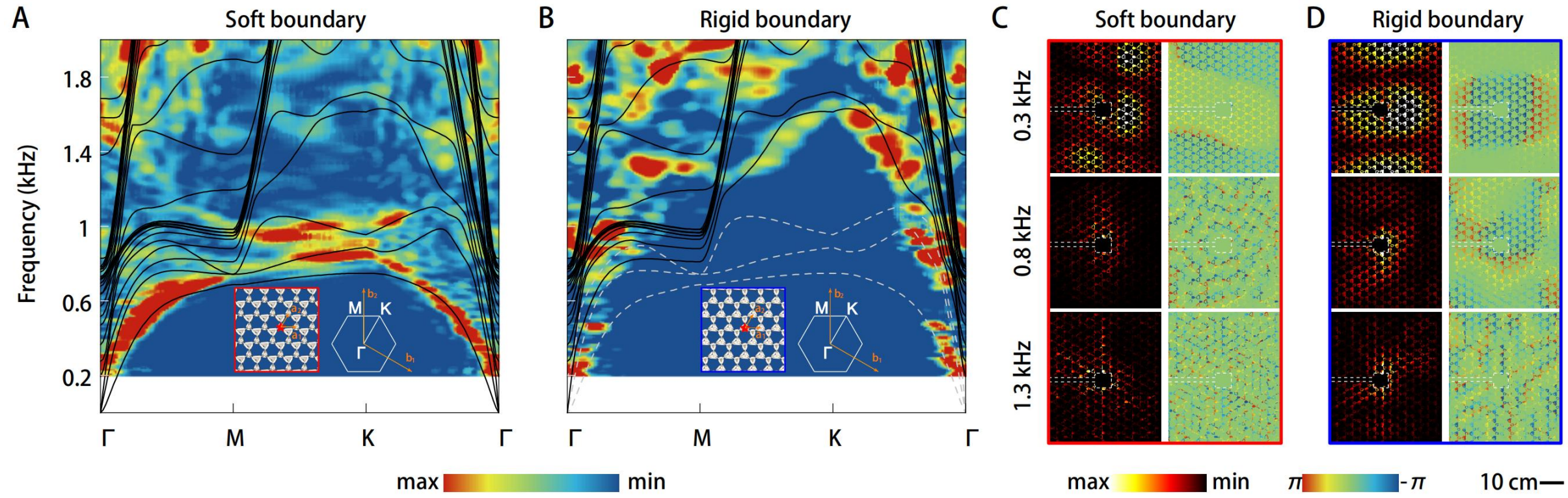

**FIG. 3. Dynamical Elasticity of the Topologically Polarized Metamaterial.** (**A**) Band structure measured at the soft boundary. (**B**) Band structure measured at the rigid boundary. The black solid curves denote the dispersion relations of bulk and topological boundary phononic modes, while the white dashed curves highlight the absence of topological edge soft modes when measured from the rigid boundary. In the insets, the left panels present 2D top views of the metamaterial, viewed from the soft boundary in (**A**) and from the rigid boundary in (**B**). The red star marks the point-shaking position, and the orange arrows indicate the primitive vectors $\boldsymbol{a}_1$ and $\boldsymbol{a}_2$ of the structure. The right panels show the two-dimensional surface Brillouin zone, with high-symmetry points labeled and orange arrows denoting the reciprocal vectors $\boldsymbol{b}_1$ and $\boldsymbol{b}_2$. (**C**) (**D**) Experimental measurements of the surface deformation profiles under excitation frequencies f=0.3 kHz, 0.8 kHz, and 1.3 kHz at the soft (**C**) and rigid (**D**) boundaries. For each case, left panels show surface amplitude distributions; right panels show corresponding phase profiles of the phononic waves.

The amplitude and phase distributions at soft and rigid boundaries are shown in Figs. 3(**C**) and 3(**D**) for frequencies of 0.3 kHz, 0.8 kHz, and 1.3 kHz, respectively. At 0.3 kHz, the soft boundary excites both edge and bulk modes, whereas the rigid boundary excites only bulk modes, as shown in Figs. 3(**C**) and (**D**). Because the penetration depth of the topological edge modes exceeds the sample size, their spatial profiles appear indistinguishable from bulk modes, leading to comparable amplitudes and wavenumbers for both boundaries. This finite-size effect defines the lower bound of the "low-frequency" working regime of our acoustic metamaterial. At 0.8 kHz, the phase distributions show that the soft boundary supports wavelengths comparable to the unit-cell scale, consistent with the dispersion of topological surface modes in Fig. 3(**A**). Conversely, the rigid boundary exhibits longer wavelengths in the small-wavenumber regime of the bulk band structure, as shown in Fig. 3(**B**). Although the surface responses at 0.8 kHz display markedly different wavelengths for opposing surfaces, as shown for 0.8 kHz in Figs. 3(**C**) and (**D**), their mode amplitudes remain comparable because the soft boundary excites both edge and bulk modes, whereas the rigid boundary excites bulk modes that also contribute to the observed amplitudes. At higher frequencies (1.3 kHz), bulk modes dominate for both boundary excitation inputs, resulting in nearly identical amplitude and phase distributions due to their mechanically similar responses. At 1.3 kHz, the maximum frequency of the topological edge modes, this value defines the upper limit of the metamaterial's low frequency operating regime.

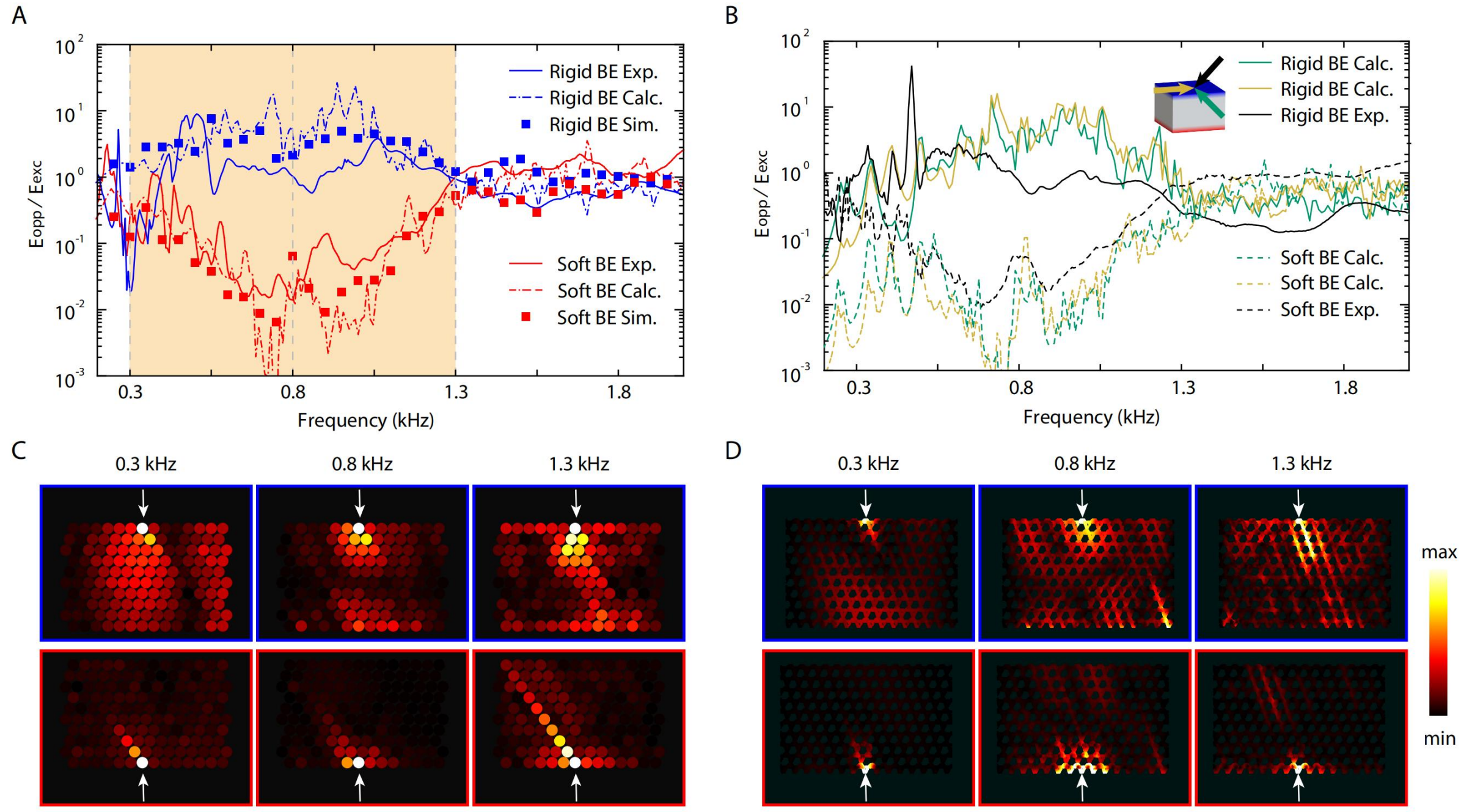

**FIG. 4. Omnidirectional and Asymmetric Wave Isolation in Topologically Polarized Elastic Metamaterials.** (**A**) The ratio of output kinetic energy under normal inward boundary excitation (denoted as BE in the inset), measured at both soft and rigid surfaces. Solid lines represent experimental measurements, dashed lines indicate theoretical predictions, and symbol □ corresponds to numerical simulations. (**B**) The ratio of output kinetic energy at soft and rigid surfaces under three distinct boundary excitations: two orthogonal in-plane forces parallel to the surface boundaries (green and yellow curves, measured via numerical simulation), and a force applied at a 45° angle relative to the surface normal (black curve, measured experimentally), revealing omnidirectional asymmetry in elastic wave response. (**C**) Amplitude spatial profiles of the experimentally measured elastic wave propagation across the metamaterial at excitation frequencies of 0.3, 0.8, and 1.3 kHz, demonstrating pronounced asymmetric characteristics in the boundary-localized kinetic energy. (**D**) Simulated amplitude spatial profiles of elastic wave propagation at excitation frequencies of 0.3, 0.8, and 1.3 kHz, corresponding to the boundary excitation described in (**A**), further confirming asymmetric localization of the boundary kinetic energy across the metamaterial.

To quantify the observed asymmetry of wave isolation, we define the metric as the ratio of output kinetic energy between the excitation and opposite boundaries. In the experiment, a chirp signal with scanning frequency from $f_{\mathrm{min}} = 0.2$ kHz to $f_{\mathrm{max}} = 2$ kHz and time window of 10s is imposed through a shaker at the center of the input boundary, controlling the displacement $\boldsymbol{x}(\boldsymbol{r}_{\mathrm{center}}, t)$. The vibration velocity $\boldsymbol{v}(\boldsymbol{r}, f)$ is recorded by a network analyzer across the input boundary for $f_{\mathrm{min}} \leq f \leq f_{\mathrm{max}}$, with successive sites measured at intervals of $d = 2.5$ mm. On the opposite boundary, velocity measurements are taken at identical spatial intervals within the same frequency range, yielding approximately 8000 sampling points on each boundary. For each frequency , we compute the squared velocity and average over all sampling points to obtain the mean kinetic energy on the input and output boundaries: $\langle v_i^2(f)\rangle = \frac{1}{N_{\mathrm{i}}}\sum_{j=1}^{N_{\mathrm{i}}} v^2(\boldsymbol{r}_{\mathrm{j}} \in \text{ measuring surface}, f)$ , where $i = \mathrm{exc}$ denotes the excitation boundary and $i = \mathrm{opp}$ the opposite boundary, and $N_{\mathrm{i}}$ denotes the sampling points. The ratio of output kinetic energy between the opposite boundaries is then defined as

$$T(f) = \langle v^2{}_{\mathrm{opp}}(f)\rangle \,/\, \langle v^2{}_{\mathrm{exc}}(f)\rangle$$

Although the computed quantity is the average squared velocity rather than kinetic energy, the two are directly proportional, so this serves as a consistent metric.

As shown in Fig. 4(**A**), the kinetic-energy ratio displays pronounced asymmetry in the low-frequency regime. This range extends from 0.3 kHz, corresponding to the edge mode whose penetration depth is comparable to the system size, up to 1.3 kHz, the highest edge-mode frequency. Within this regime, soft-boundary excitation produces pronounced minima, revealing strong localization of energy at the excited surface. In contrast, rigid-boundary excitation generates distinct maxima in the same frequency range, signifying efficient bulk-mediated wave transmission and the subsequent recapture of elastic energy at the opposite soft boundary. This marked asymmetry highlights the fundamentally different penetration depths of topological edge modes and bulk modes, in agreement with the distinct surface responses shown in Fig. 3(**C**) and (**D**) at 0.8 kHz. Further analytic and numerical investigations of the kinetic-energy ratio, obtained by varying the damping coefficient, are shown in Fig. S4.

Having established asymmetric energy isolation under inward-directed normal excitation, we now investigate whether this phononic asymmetry persists under omnidirectional external forces. To address this, we examine the energy contrast induced by three distinct boundary excitations: two orthogonal in-plane forces aligned parallel to the surface boundaries, and one force applied at a 45° angle relative to the surface normal. The resulting kinetic energy ratio spectra reveal a robust omnidirectional asymmetry in the elastic wave response, as shown in Fig. 4(**B**).

To further elucidate the differences between the excitations on the soft and rigid boundaries, we numerically compute and experimentally measure the spatial distribution of wave energy across the sample's cross-section, as shown in Figs. 4(**C**) and 4(**D**), respectively. Intriguingly, while the boundary wave profiles at 0.3 kHz appear similar in Figs. 3(**C**, **D**) and in Fig. 4(**A**), the cross-sectional distributions at 0.3 kHz and 0.8 kHz clearly distinguish bulk and surface modes. At 1.3 kHz, both excitation conditions couple with bulk acoustic modes, enabling direct wave transmission through the metamaterial. In addition, we performed impact experiments on both boundaries to probe the macroscopic and transient dynamics of our three-dimensional topological elastic metamaterial (see Fig. S5, Supplementary Materials), with the experimental setup detailed in Fig. S10.

## Discussion

In this work, we present the first 3D topologically polarized elastic metamaterial that achieves omnidirectional, asymmetric wave energy isolation in the low-frequency regime, spanning from the highest edge-mode frequency down to the lower bound imposed by the finite-size effect of the sample dimensions. By designing finite-thickness ligaments, we achieve enhanced mechanical stability compared to idealized isostatic structures while maintaining robust topological protection, evidenced by a pronounced stiffness contrast between paired and parallel open surfaces. A key advance is the realization of a bandgap with a zero-frequency lower bound without a substrate, eliminating the sub-gap conduction bands characteristic of conventional phononic metamaterials. Dynamically, these metamaterials support topological soft modes that localize exclusively at a single boundary. The finite hinge thickness shifts these zero-frequency modes to finite frequencies with non-zero group velocity, enabling low-frequency vibration isolation that effectively confines energy at soft surfaces.

Our framework establishes a transformative paradigm for controlling long-wavelength classical waves, including strongly asymmetric non-local resonances. Future research directions include: exploring elastic non-reciprocity through nonlinear topological mechanics (*38-41*), investigating non-Hermitian dynamics via active components (*42-47*), and developing micro/nano strategies for topologically polarized soft matter (*48, 49*).

## Materials and Methods

### Sample preparation

The samples are fabricated by 3D printing technology, employing photosensitive resin molding. In the sample, cylinders are used to connect the lattice sites, simulating nearest-neighbor connections, while we designed flexible hinges to replace spherical hinges, which can provide bending stiffness. The thinnest part of the hinge has a diameter of $d$ = 1.5 mm, and its shape is obtained by revolved cutting with an elliptical cross-section. We ultimately constructed a periodic structure of size 15×15×10, corresponding to the lattice vectors $\boldsymbol{a}_1$, $\boldsymbol{a}_2$, and $\boldsymbol{a}_3$, respectively. An outer frame was added around the sample to serve as fixed boundary condition. The final dimensions of the sample are 521mm×470mm×272mm (Supplementary Material VI).

**Experimental measurement**

For the static measurements, we characterize the sample's response under quasistatic loading. We fixed the force sensor (HYLY-019) onto the displacement platform and brought it into contact with the sample. By slowly moving the displacement platform, we applied quasi-static pressure to the sample. The real-time displacement of the excited point was simultaneously measured using a laser vibrometer (Polytec OFV-5000), and both the deformation data and the force sensor signals were synchronously transmitted to the oscilloscope (KEYSIGHT DSOX4054A). The deformation of the sample surfaces was measured by a laser range finder (BOJKE BLG-85N-485) (Supplementary Material (*43*) IV.A).

For the dynamic measurements, we use vibration excitation source (P-820 piezoelectric actuator from PI company) to excite the sample, and then to measure its responses. We used a network analyzer (Keysight E5061B) to measure the sample's vibration combined with the laser vibrometer. Measurements were conducted on both the top and bottom surfaces, as well as the interior of the sample. These signals are fed back to the network analyzer to obtain amplitude and phase information. By performing a two-dimensional Fourier transform, we are able to get the band structure of the dispersion. And we can get the ratio of kinetic energy on opposite boundaries of the sample by amplitude information (Supplementary Material[32] IV.B).

For the impact measurements, we mounted the high-speed force sensor (PCB-208C01) on a linear guide rail. An electromagnetic switch was used to release it, allowing for vertical impact from the same release height on both the top and bottom surfaces of the sample. The real-time signal was acquired via an oscilloscope (KEYSIGHT DSOX4054A) to obtain the impact force-time curve (Supplementary Material VI.C) .

**Numerical simulation**

We calculate the dynamical responses of the deformed pyrochlore lattice with the spring-mass model by Linear Response function. By separating the displacement vector into the boundary-driven component (i.e., $u_c$) and the responding component ($u_f$), which leads to the block-matrix formulation of the Newtonian dynamics, and for specific excitation with frequency $\omega$, we can get the corresponding Green's function to calculate the dynamical responses (detailed in the Supplementary Material II).

The Solid Mechanics Module in COMSOL Multiphysics was employed to simulate the continuous model used in the experiment. The Eigen-frequency study is performed on a supercell composed of 10 unit cells to derive the phonon dispersion band structure. For the analysis of both static and dynamic responses, the same model as that of the experimental specimen is employed. Fixed boundary conditions are applied around the sample. The Stationary study is conducted for the static analysis, while a frequency-domain study is utilized for the dynamic analysis. In the static simulation, a force boundary condition is imposed at the excitation site, whereas in the dynamic simulation, a displacement boundary condition is applied, ensuring consistency with the experimental setup.

**Data availability**

This study did not generate new materials. All data and code needed to evaluate and reproduce the results in the paper are present in the paper and/or the Supplementary Materials. The data supporting the plots in this paper were produced using MATLAB. All MATLAB codes have been made freely accessible at https://doi.org/10.5281/zenodo.18707973.

**Acknowledgements**

This work was supported by the National Nature Science Foundation of China under award number Nos. 12374157, 12272040, and 12302112 (D.Z. F.L., and Y.W.), and the Fundamental Research Funds for the Central Universities (No. 30923010207).

**Author contributions**

Conceptualization: FL, DZ. Experiment: FL, DZ, SZ. Modelling: XG, FM. Writing: SZ, XG, FM, ZT, YW, DZ, FL, YY.

**Competing interests**

The authors declare they have no competing interest.

# 3D Topologically Polarized Elastic Metamaterials Enable Asymmetric Energy Isolation at Finite Frequencies: Supplementary Materials

Shaoyuan Zhang,[1,*] Xuejian Gong,[1,*] Fangyuan Ma,[1] Zheng Tang,[1] Ying Wu,[2,†] Di Zhou,[1,‡] Feng Li,[1,§] and Yugui Yao[1]

[1]*Key Lab of Advanced Optoelectronic Quantum Architecture and Measurement (MOE), School of Physics, Beijing Institute of Technology, Beijing, 100081, China*

[2]*School of Physics, Nanjing University of Science and Technology, Nanjing 210094, China*

## I. TOPOLOGICAL POLARIZATION AND MECHANICAL PROPERTIES OF THE SOFT METAMATERIAL

This section focuses on constructing the topological mechanical phase of the 3D-printed soft metamaterial. As discussed in the main text, the topological nature of the soft phononic modes is derived from the topological polarization of the mechanical properties in the theoretical spring-mass model. The unit cell of this model consists of mass particles connected by nearest-neighbor and next-nearest-neighbor Hookean springs, as illustrated in SI. Fig. S1. Below, we construct the lattice Hamiltonian and the topological polarization in this deformed pyrochlore lattice.

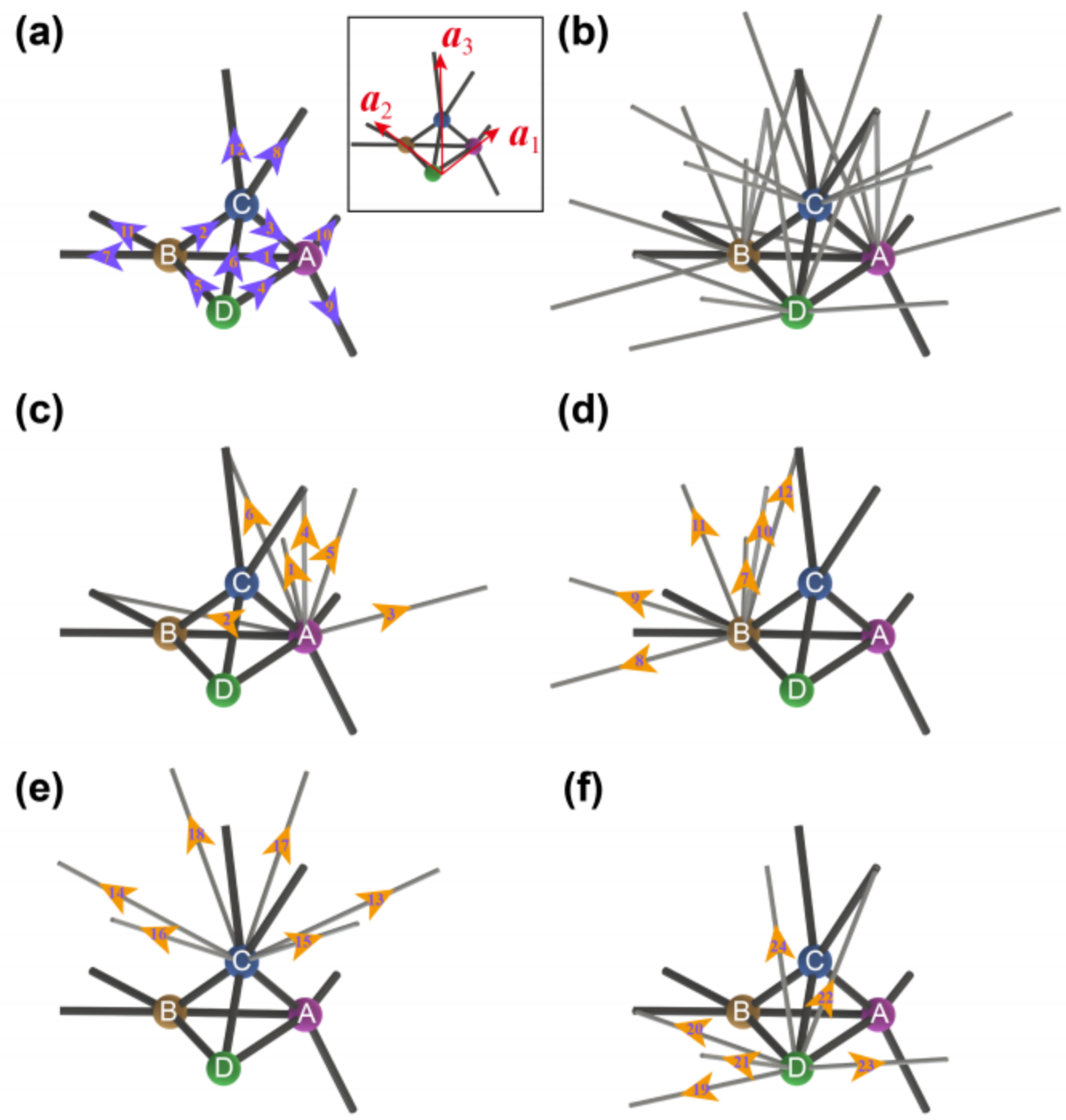

FIG. S1. The geometric configuration of the unit cell in the deformed pyrochlore lattice includes nodes, solid black lines, and solid gray lines, which respectively represent the sites, nearest-neighbor bonds, and next-nearest-neighbor bonds within the unit cell. (a) Unit cell with NN bonds. (b) unit cell with NNN bonds. (c)-(f) All NNN bonds for A, B, C and D sites, respectively. NN and NNN bonds within the lattice are labeled 1–12 and 1–24, with their orientations indicated by purple and orange arrows, respectively.

In the three-dimensional lattice, the unit cell is labeled by three integer numbers, $n_1$, $n_2$, and $n_3$, indicating its position at $n_1\boldsymbol{a}_1 + n_2\boldsymbol{a}_2 + n_3\boldsymbol{a}_3$. Throughout this Supplementary Material, a vectorial index $\boldsymbol{n} = (n_1, n_2, n_3)$ is used to denote the unit cell. Following this convention, the sites are labeled as $A(\boldsymbol{n})$, $B(\boldsymbol{n})$, $C(\boldsymbol{n})$, and $D(\boldsymbol{n})$, with their positions given by $X(\boldsymbol{n}) = X + n_1\boldsymbol{a}_1 + n_2\boldsymbol{a}_2 + n_3\boldsymbol{a}_3$ for $X = A, B, C, D$.

* These authors contribute equally to this work.
† yingwu@njust.edu.cn
‡ dizhou@bit.edu.cn
§ phlifeng@bit.edu.cn

Given the site positions detailed in the main text, we discuss the connectivity of nodes within the deformed pyrochlore lattice network. This requires a clear definition of the 12 nearest-neighbor edges and the 24 next-nearest-neighbor edges within the unit cell of the spring-mass model, as illustrated in SI. Fig. S1. These nearest-neighbor edge vectors are defined as:

$$\boldsymbol{l}_i(\boldsymbol{n}) = \boldsymbol{l}_i = \overline{X(\boldsymbol{n})Y(\boldsymbol{m})} = \boldsymbol{Y}(\boldsymbol{m}) - \boldsymbol{X}(\boldsymbol{n}), \tag{1}$$

with $i = 1, 2, \ldots, 12$. Here, the equality $\boldsymbol{l}_i(\boldsymbol{n}) = \boldsymbol{l}_i$ arises from the discrete translational symmetry of the lattice. Specifically, we have $\boldsymbol{l}_1 = \overline{A(\boldsymbol{n})B(\boldsymbol{n})}$, $\boldsymbol{l}_2 = \overline{B(\boldsymbol{n})C(\boldsymbol{n})}$, $\boldsymbol{l}_3 = \overline{C(\boldsymbol{n})A(\boldsymbol{n})}$, $\boldsymbol{l}_4 = \overline{D(\boldsymbol{n})A(\boldsymbol{n})}$, $\boldsymbol{l}_5 = \overline{D(\boldsymbol{n})B(\boldsymbol{n})}$, $\boldsymbol{l}_6 = \overline{D(\boldsymbol{n})C(\boldsymbol{n})}$, $\boldsymbol{l}_7 = \overline{B(\boldsymbol{n})A(\boldsymbol{n}+(-1,1,0))}$, $\boldsymbol{l}_8 = \overline{C(\boldsymbol{n})B(\boldsymbol{n}+(0,-1,1))}$, $\boldsymbol{l}_9 = \overline{A(\boldsymbol{n})C(\boldsymbol{n}+(1,0,-1))}$, $\boldsymbol{l}_{10} = \overline{A(\boldsymbol{n})D(\boldsymbol{n}+(1,0,0))}$, $\boldsymbol{l}_{11} = \overline{B(\boldsymbol{n})D(\boldsymbol{n}+(0,1,0))}$, and $\boldsymbol{l}_{12} = \overline{C(\boldsymbol{n})D(\boldsymbol{n}+(0,0,1))}$. The change in edge length is given by:

$$\delta l_i(\boldsymbol{n}) = [\boldsymbol{u}_Y(\boldsymbol{m}) - \boldsymbol{u}_X(\boldsymbol{n})] \cdot \hat{t}_i \tag{2}$$

The edge orientations are defined as

$$\hat{t}_i = \boldsymbol{l}_i/|\boldsymbol{l}_i|, \qquad i = 1, 2, \ldots, 12. \tag{3}$$

Spatially repetitive frames enable the representation of phonon modes as plane waves. These modes can be mathematically formulated using the Fourier transformation: $\boldsymbol{u}_X(\boldsymbol{n}) = \sum_{\boldsymbol{k}} \boldsymbol{u}_X(\boldsymbol{k}) e^{\mathrm{i}\boldsymbol{k}\cdot\boldsymbol{r}(\boldsymbol{n})}$. In this equation, $\boldsymbol{u}_X(\boldsymbol{n})$ represents the mechanical wave for the $X = A, B, C, D$ site in the unit cell, labeled by the vector index $\boldsymbol{n} = (n_1, n_2, n_3)$. The corresponding position is given by $\boldsymbol{r}(\boldsymbol{n}) = n_1\boldsymbol{a}_1 + n_2\boldsymbol{a}_2 + n_3\boldsymbol{a}_3$, with $\boldsymbol{a}_1$, $\boldsymbol{a}_2$, and $\boldsymbol{a}_3$ the primitive vectors. $\boldsymbol{k}$ represents the phonon wavevector, and the summation $\sum_{\boldsymbol{k}}$ runs over the three-dimensional Brillouin zone. Furthermore, $\boldsymbol{u}_X(\boldsymbol{k})$ denotes the phonon mode at the wavevector $\boldsymbol{k}$, with $X = A, B, C, D$.

We denote $\boldsymbol{u}(\boldsymbol{k}) = (\boldsymbol{u}_A(\boldsymbol{k}), \boldsymbol{u}_B(\boldsymbol{k}), \boldsymbol{u}_C(\boldsymbol{k}), \boldsymbol{u}_D(\boldsymbol{k}))^\top$ as the $12\times1$ displacement vector for the momentum $\boldsymbol{k}$. Similarly, bond elongations can be decomposed into plane-wave format as $\delta l_i(\boldsymbol{n}) = \sum_{\boldsymbol{k}} \delta l_i(\boldsymbol{k}) e^{\mathrm{i}\boldsymbol{k}\cdot\boldsymbol{r}(\boldsymbol{n})}$ for $i = 1, 2, \ldots, 12$, where $\delta l_i(\boldsymbol{k})$ represents the Fourier-transformed bond elongations at the momentum $\boldsymbol{k}$. Consequently, we define $\delta\boldsymbol{l}(\boldsymbol{k}) = (\delta l_1(\boldsymbol{k}), \delta l_2(\boldsymbol{k}), \ldots, \delta l_{12}(\boldsymbol{k}))^\top$ as the $12 \times 1$ column vector for the bond elongations of wavevector $\boldsymbol{k}$. These Fourier-transformed quantities are related by the Fourier-transformed $12\times12$ compatibility matrix $\mathbf{C}(\boldsymbol{k})$ as $\mathbf{C}(\boldsymbol{k})\boldsymbol{u}(\boldsymbol{k}) = \delta\boldsymbol{l}(\boldsymbol{k})$. For the generalized pyrochlore lattice, the Fourier-transformed compatibility matrix is detailed as follows,

$$\mathbf{C}(\boldsymbol{k}) = \begin{pmatrix} -\hat{t}_1 & \hat{t}_1 & 0 & 0 \\ 0 & -\hat{t}_2 & \hat{t}_2 & 0 \\ \hat{t}_3 & 0 & -\hat{t}_3 & 0 \\ \hat{t}_4 & 0 & 0 & -\hat{t}_4 \\ 0 & \hat{t}_5 & 0 & -\hat{t}_5 \\ 0 & 0 & \hat{t}_6 & -\hat{t}_6 \\ e^{\mathrm{i}\boldsymbol{k}\cdot(\boldsymbol{a}_2-\boldsymbol{a}_1)}\hat{t}_7 & -\hat{t}_7 & 0 & 0 \\ 0 & e^{\mathrm{i}\boldsymbol{k}\cdot(\boldsymbol{a}_3-\boldsymbol{a}_2)}\hat{t}_8 & -\hat{t}_8 & 0 \\ -\hat{t}_9 & 0 & e^{\mathrm{i}\boldsymbol{k}\cdot(\boldsymbol{a}_1-\boldsymbol{a}_3)}\hat{t}_9 & 0 \\ -\hat{t}_{10} & 0 & 0 & e^{\mathrm{i}\boldsymbol{k}\cdot\boldsymbol{a}_1}\hat{t}_{10} \\ 0 & -\hat{t}_{11} & 0 & e^{\mathrm{i}\boldsymbol{k}\cdot\boldsymbol{a}_2}\hat{t}_{11} \\ 0 & 0 & -\hat{t}_{12} & e^{\mathrm{i}\boldsymbol{k}\cdot\boldsymbol{a}_3}\hat{t}_{12} \end{pmatrix}. \tag{4}$$

Similarly, the edge vectors for the next-nearest-neighbor bonds are defined as follows:

$$\boldsymbol{l}'_i(\boldsymbol{n}) = \boldsymbol{l}'_i = \overline{X'(\boldsymbol{n})Y'(\boldsymbol{m})} = \boldsymbol{Y}'(\boldsymbol{m}) - \boldsymbol{X}'(\boldsymbol{n}), \tag{5}$$

with $i = 1, 2, \ldots, 24$, as there are a total of 24 next-nearest-neighbor connections in the unit cell. Specifically, the next-nearest-neighbor bonds are classified into four types. The next-nearest-neighbor bonds starting from the $A$ site are given by: $\boldsymbol{l}'_1 = \overline{A(\boldsymbol{n})B(\boldsymbol{n}+(1,0,0))}$, $\boldsymbol{l}'_2 = \overline{A(\boldsymbol{n})D(\boldsymbol{n}+(0,1,0))}$, $\boldsymbol{l}'_3 = \overline{A(\boldsymbol{n})C(\boldsymbol{n}+(1,-1,0))}$, $\boldsymbol{l}'_4 = \overline{A(\boldsymbol{n})B(\boldsymbol{n}+(0,-1,1))}$, $\boldsymbol{l}'_5 = \overline{A(\boldsymbol{n})C(\boldsymbol{n}+(1,0,0))}$, $\boldsymbol{l}'_6 = \overline{A(\boldsymbol{n})D(\boldsymbol{n}+(0,0,1))}$. The next-nearest-neighbor bonds that are connected to the $B$-sites: $\boldsymbol{l}'_7 = \overline{B(\boldsymbol{n})A(\boldsymbol{n}+(0,1,0))}$, $\boldsymbol{l}'_8 = \overline{B(\boldsymbol{n})D(\boldsymbol{n}+(-1,1,0))}$, $\boldsymbol{l}'_9 = \overline{B(\boldsymbol{n})C(\boldsymbol{n}+(-1,1,0))}$, $\boldsymbol{l}'_{10} = \overline{B(\boldsymbol{n})A(\boldsymbol{n}+(-1,0,1))}$, $\boldsymbol{l}'_{11} = \overline{B(\boldsymbol{n})C(\boldsymbol{n}+(0,1,0))}$, $\boldsymbol{l}'_{12} = \overline{B(\boldsymbol{n})D(\boldsymbol{n}+(0,0,1))}$. The next-nearest-neighbor bonds that are connected to the $C$-sites: $\boldsymbol{l}'_{13} = \overline{C(\boldsymbol{n})A(\boldsymbol{n}+(0,-1,1))}$, $\boldsymbol{l}'_{14} = \overline{C(\boldsymbol{n})B(\boldsymbol{n}+(-1,0,1))}$, $\boldsymbol{l}'_{15} =$ $\overline{C(\boldsymbol{n})D(\boldsymbol{n}+(0,-1,1))}$, $\boldsymbol{l}'_{16} = \overline{C(\boldsymbol{n})D(\boldsymbol{n}+(-1,0,1))}$, $\boldsymbol{l}'_{17} = \overline{C(\boldsymbol{n})A(\boldsymbol{n}+(0,0,1))}$, $\boldsymbol{l}'_{18} = \overline{C(\boldsymbol{n})B(\boldsymbol{n}+(0,0,1))}$. The next-nearest-neighbor bonds that are connected to the $D$-sites: $\boldsymbol{l}'_{19} = \overline{D(\boldsymbol{n})B(\boldsymbol{n}+(-1,0,0))}$, $\boldsymbol{l}'_{20} = \overline{D(\boldsymbol{n})A(\boldsymbol{n}+(-1,1,0))}$, $\boldsymbol{l}'_{21} = \overline{D(\boldsymbol{n})C(\boldsymbol{n}+(-1,0,0))}$, $\boldsymbol{l}'_{22} = \overline{D(\boldsymbol{n})B(\boldsymbol{n}+(0,-1,1))}$, $\boldsymbol{l}'_{23} = \overline{D(\boldsymbol{n})C(\boldsymbol{n}+(0,-1,0))}$, $\boldsymbol{l}'_{24} = \overline{D(\boldsymbol{n})A(\boldsymbol{n}+(-1,0,1))}$. The change in the edge length reads

$$\delta l'_i(\boldsymbol{n}) = [\boldsymbol{u}_{Y'}(\boldsymbol{m}) - \boldsymbol{u}_{X'}(\boldsymbol{n})] \cdot \hat{t}'_i \tag{6}$$

where the unit vector for the bond orientation reads

$$\hat{t}'_i = \boldsymbol{l}'_i/|\boldsymbol{l}'_i|, \qquad i = 1, 2, \cdots, 24 \tag{7}$$

The compatibility matrix of the next-nearest-neighbor bonds is a $24 \times 12$ matrix, which is given by

$$\mathbf{C}'(\boldsymbol{k}) = \begin{pmatrix}
-\hat{t}'_1 & e^{\mathrm{i}\boldsymbol{k}\cdot\boldsymbol{a}_1}\hat{t}'_1 & 0 & 0 \\
-\hat{t}'_2 & 0 & 0 & e^{\mathrm{i}\boldsymbol{k}\cdot\boldsymbol{a}_2}\hat{t}'_2 \\
-\hat{t}'_3 & 0 & e^{\mathrm{i}\boldsymbol{k}\cdot(\boldsymbol{a}_1-\boldsymbol{a}_2)}\hat{t}'_3 & 0 \\
-\hat{t}'_4 & e^{\mathrm{i}\boldsymbol{k}\cdot(\boldsymbol{a}_3-\boldsymbol{a}_2)}\hat{t}'_4 & 0 & 0 \\
-\hat{t}'_5 & 0 & e^{\mathrm{i}\boldsymbol{k}\cdot\boldsymbol{a}_1}\hat{t}'_5 & 0 \\
-\hat{t}'_6 & 0 & 0 & e^{\mathrm{i}\boldsymbol{k}\cdot\boldsymbol{a}_3}\hat{t}'_6 \\
e^{\mathrm{i}\boldsymbol{k}\cdot\boldsymbol{a}_2}\hat{t}'_7 & -\hat{t}'_7 & 0 & 0 \\
0 & -\hat{t}'_8 & 0 & e^{\mathrm{i}\boldsymbol{k}\cdot(\boldsymbol{a}_2-\boldsymbol{a}_1)}\hat{t}'_8 \\
0 & -\hat{t}'_9 & e^{\mathrm{i}\boldsymbol{k}\cdot(\boldsymbol{a}_2-\boldsymbol{a}_1)}\hat{t}'_9 & 0 \\
e^{\mathrm{i}\boldsymbol{k}\cdot(\boldsymbol{a}_3-\boldsymbol{a}_1)}\hat{t}'_8 & -\hat{t}'_{10} & 0 & 0 \\
0 & -\hat{t}'_{11} & e^{\mathrm{i}\boldsymbol{k}\cdot\boldsymbol{a}_2}\hat{t}'_{11} & 0 \\
0 & -\hat{t}'_{12} & 0 & e^{i\boldsymbol{k}\cdot\boldsymbol{a}_3}\hat{t}'_{12} \\
e^{\mathrm{i}\boldsymbol{k}\cdot(\boldsymbol{a}_3-\boldsymbol{a}_2)}\hat{t}'_{13} & 0 & -\hat{t}'_{13} & 0 \\
0 & e^{i\boldsymbol{k}\cdot(\boldsymbol{a}_3-\boldsymbol{a}_1)}\hat{t}'_{14} & -\hat{t}'_{14} & 0 \\
0 & 0 & -\hat{t}'_{15} & e^{\mathrm{i}\boldsymbol{k}\cdot(\boldsymbol{a}_3-\boldsymbol{a}_2)}\hat{t}'_{15} \\
0 & 0 & -\hat{t}'_{16} & e^{\mathrm{i}\boldsymbol{k}\cdot(\boldsymbol{a}_3-\boldsymbol{a}_1)}\hat{t}'_{16} \\
e^{\mathrm{i}\boldsymbol{k}\cdot\boldsymbol{a}_3}\hat{t}'_{17} & 0 & -\hat{t}'_{17} & 0 \\
0 & e^{\mathrm{i}\boldsymbol{k}\cdot\boldsymbol{a}_3}\hat{t}'_{18} & -\hat{t}'_{18} & 0 \\
0 & e^{-\mathrm{i}\boldsymbol{k}\cdot\boldsymbol{a}_1}\hat{t}'_{19} & 0 & -\hat{t}'_{19} \\
e^{\mathrm{i}\boldsymbol{k}\cdot(\boldsymbol{a}_2-\boldsymbol{a}_1)}\hat{t}'_{20} & 0 & 0 & -\hat{t}'_{20} \\
0 & 0 & e^{-\mathrm{i}\boldsymbol{k}\cdot\boldsymbol{a}_1}\hat{t}'_{21} & -\hat{t}'_{21} \\
0 & e^{\mathrm{i}\boldsymbol{k}\cdot(\boldsymbol{a}_3-\boldsymbol{a}_2)}\hat{t}'_{22} & 0 & -\hat{t}'_{22} \\
0 & 0 & e^{-\mathrm{i}\boldsymbol{k}\cdot\boldsymbol{a}_2}\hat{t}'_{23} & -\hat{t}'_{23} \\
e^{\mathrm{i}\boldsymbol{k}\cdot(\boldsymbol{a}_3-\boldsymbol{a}_1)}\hat{t}'_{24} & 0 & 0 & -\hat{t}'_{24}
\end{pmatrix}. \tag{8}$$

Based on the definition of the compatibility matrices for the nearest-neighbor and next-nearest-neighbor bonds, we now consider the Newtonian dynamics $m\omega^2(\boldsymbol{k})\boldsymbol{u}(\boldsymbol{k}) = \mathbf{D}(\boldsymbol{k})\boldsymbol{u}(\boldsymbol{k})$, where the dynamical matrix $\mathbf{D}(\boldsymbol{k})$ is given by the compatibility matrices as follows,

$$\mathbf{D}(\boldsymbol{k}) = k\mathbf{C}^\dagger(\boldsymbol{k})\mathbf{C}(\boldsymbol{k}) + \kappa\mathbf{C}'^\dagger(\boldsymbol{k})\mathbf{C}'(\boldsymbol{k}). \tag{9}$$

Here, $k$ and $\kappa$ denote the elastic stiffness for the nearest-neighbor and next-nearest-neighbor bonds, respectively.

In the limit when next-nearest-neighbor interactions are negligible, i.e., $\kappa \to 0$, the lattice dynamics is dominated by springs of nearest-neighbor interactions. Every unit cell contains four sites, resulting in a total of $4d = 12$ degrees of freedom ($d = 3$ is the dimensionality). On the other hand, the unit cell contains 12 constraints, leaving the deformed pyrochlore lattice have the balanced degrees of freedom and constraints, and is therefore isostatic. The topological properties of the static mechanics in this isostatic structure (i.e., by taking the limit of $\kappa \to 0$) modes can be captured by the integer-valued winding numbers defined from the compatibility matrix in the reciprocal space. Given the wavevector $\boldsymbol{k}$ in the three-dimensional Brillouin zone, topological winding numbers are defined as follows,

$$\mathcal{N}_i(\boldsymbol{k}) = -\frac{1}{2\pi\mathrm{i}}\oint_{\boldsymbol{k}\to\boldsymbol{k}+\boldsymbol{b}_i} d\boldsymbol{k} \cdot \nabla_{\boldsymbol{k}} \ln\det\mathbf{C}(\boldsymbol{k}), \tag{10}$$

where $i = 1, 2, 3$ labels the three reciprocal vector directions, $\boldsymbol{b}_i$ is the $i$-th reciprocal vector, and the integration trajectory, $\boldsymbol{k} \to \boldsymbol{k} + \boldsymbol{b}_i$, follows a straight and closed loop that is parallel to $\boldsymbol{b}_i$. For each $\boldsymbol{k}$, there are a total of three topological winding numbers in the generalized pyrochlore lattice.

For certain geometric configurations of the generalized pyrochlore lattice, such as the unit cell configuration displayed in Fig. 1**b** of the main text, the phonon band structure can be gapped throughout the three-dimensional Brillouin zone, except for the trivial $\boldsymbol{k}=0$ point. This $\boldsymbol{k}=0$ point represents the wavevector of mechanical zero modes that translate or rotate the entire lattice as a whole, commonly referred to as "trivial mechanical zero modes" (*44*). This gapped mechanical phase can be realized in the deformed pyrochlore lattice. As a result, all winding numbers, i.e., $(\mathcal{N}_1(\boldsymbol{k}), \mathcal{N}_2(\boldsymbol{k}), \mathcal{N}_3(\boldsymbol{k})) = (1,-1,-2)$, remain invariant for arbitrary $\boldsymbol{k}$ in 3D reciprocal space.

Furthermore, these *invariant* integer numbers can be combined into a topologically invariant vector, called the topological polarization,

$$\boldsymbol{R}_{\mathrm{T}} = \sum_{i=1}^{3} \mathcal{N}_i \boldsymbol{a}_i. \tag{11}$$

For the geometric configuration presented in this work (Fig. 2**a** of the main text), the topological polarization reads $\boldsymbol{R}_{\mathrm{T}} = \boldsymbol{a}_1 - \boldsymbol{a}_2 - 2\boldsymbol{a}_3$. The physical significance of this topological polarization is that the topological floppy modes localize preferably on the surface that the topological polarization points to, whereas the parallel opposite surface has fewer topological mechanical floppy modes. In other words, $\boldsymbol{R}_{\mathrm{T}}$ indicates the topological polarization of the boundary floppy modes in the gapped phase of the considered isostatic lattice.

The principle of bulk-boundary correspondence states that the topological polarization obtained from the bulk phonon bands plays a crucial role in determining the number density of mechanical floppy modes on open surfaces of isostatic lattices. However, the local details of how nodes and bonds are connected on the open surfaces also influence the distribution of floppy modes. This local connectivity is quantitatively captured by the local polarization vector $\boldsymbol{R}_{\mathrm{L}}$. Together, topological and local polarizations govern the number distribution of boundary floppy modes via the relationship:

$$\nu = \frac{1}{2\pi} \boldsymbol{G} \cdot (\boldsymbol{R}_{\mathrm{T}} + \boldsymbol{R}_{\mathrm{L}}), \tag{12}$$

where the reciprocal lattice vector $\boldsymbol{G}$ is perpendicular to the considered open surface and points outward. In Fig. 1**d** and Fig. 2**a**, the reciprocal vectors for the top and bottom boundaries, perpendicular to $\boldsymbol{b}_3$, are $\boldsymbol{G}_{\uparrow} = +\boldsymbol{b}_3$ and $\boldsymbol{G}_{\downarrow} = -\boldsymbol{b}_3$, respectively. The local polarizations are $\boldsymbol{R}_{\mathrm{L}\uparrow} = -\boldsymbol{a}_2 + 2\boldsymbol{a}_3$ and $\boldsymbol{R}_{\mathrm{L}\downarrow} = \boldsymbol{a}_1 - \boldsymbol{a}_3$ for the top and bottom open surfaces, respectively. Consequently, the number densities of surface floppy modes per supercell are $\nu_{\uparrow} = \boldsymbol{b}_3 \cdot (\boldsymbol{R}_{\mathrm{T}} + \boldsymbol{R}_{\mathrm{L}\uparrow})/2\pi = 0$ and $\nu_{\downarrow} = -\boldsymbol{b}_3 \cdot (\boldsymbol{R}_{\mathrm{T}} + \boldsymbol{R}_{\mathrm{L}\downarrow})/2\pi = 3$ for the top and bottom surfaces.

The floppy mode distribution, $\nu_{\uparrow} = 0$, indicates that the top surface of the generalized pyrochlore lattice is completely devoid of mechanical floppy modes, rendering the open surface as rigid as the inner body of the lattice. In contrast, the floppy mode distribution, $\nu_{\downarrow} = 3$, exhibits a much softer local response due to the exclusive emergence of topological floppy modes on the bottom boundary.

## II. EQUIVALENCE BETWEEN BENDING STIFFNESS AND NEXT-NEAREST-NEIGHBOR INTERACTIONS

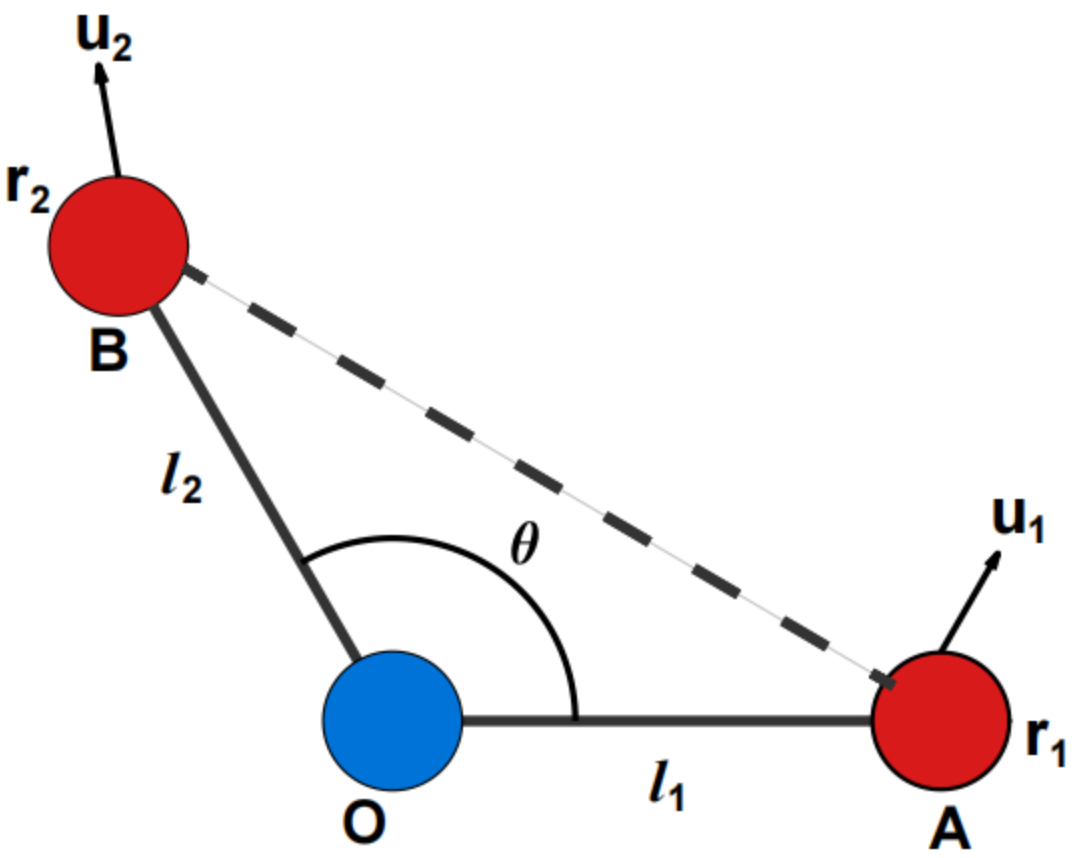

FIG. S2. Three adjacent sites in Spring-mass model.

In this section, we demonstrate that bending stiffness can be effectively modeled as a next-nearest-neighbor (NNN) interaction within the framework of linear elasticity.

Consider a central site $O$ connected to two neighboring sites $A$ and $B$, with bond lengths $OA = l_1$ and $OB = l_2$. Let the positions of sites $A$ and $B$ be denoted by $\vec{r}_1$ and $\vec{r}_2$, respectively. The rest angle at the hinge $\angle AOB$ is $\theta$, and the system is initially free of bending energy. The geometry imposes the constraint:

$$l_1^2 + l_2^2 - 2l_1 l_2 \cos\theta = |\vec{r}_2 - \vec{r}_1|^2. \tag{13}$$

Now, consider small displacements $\vec{u}_1$ and $\vec{u}_2$ applied to sites $A$ and $B$, respectively, such that $|\vec{u}_1 - \vec{u}_2| \ll |\vec{r}_1 - \vec{r}_2|$, consistent with the linear elastic regime. These displacements perturb the opening angle from $\theta$ to $\theta + \delta\theta$, where $\delta\theta \ll \theta$. The updated geometric constraint becomes:

$$l_1^2 + l_2^2 - 2l_1 l_2 \cos(\theta + \delta\theta) = \left|\vec{r}_2 + \vec{u}_2 - \vec{r}_1 - \vec{u}_1\right|^2. \tag{14}$$

Expanding both sides to first order in $\delta\theta$ and $\vec{u}_2 - \vec{u}_1$, we obtain:

$$\delta\theta = \frac{|\vec{r}_2 - \vec{r}_1|}{l_1 l_2 \sin\theta} \hat{n}_{12} \cdot (\vec{u}_2 - \vec{u}_1), \tag{15}$$

where $\hat{n}_{12} = (\vec{r}_2 - \vec{r}_1)/|\vec{r}_2 - \vec{r}_1|$ is the unit vector pointing from $A$ to $B$. The corresponding change in the bending elastic potential energy is

$$U_b = \frac{\nu}{2}\delta\theta^2 \tag{16}$$

where $\nu$ is the bending stiffness. Substituting the expression for $\delta\theta$, the bending energy can be rewritten as:

$$U_b = \frac{K'}{2}\left[\hat{n}_{12} \cdot (\vec{u}_2 - \vec{u}_1)\right]^2 = \frac{K'}{2}\delta l^2, \tag{17}$$

where $\delta l = \hat{n}_{12} \cdot (\vec{u}_2 - \vec{u}_1)$ represents the effective change in length between sites $A$ and $B$, and $K' = \nu \left|\vec{r}_2 - \vec{r}_1\right|^2 / \left(l_1^2 l_2^2 \sin^2\theta\right)$ is the effective spring constant of a Hookean spring placed directly between $A$ and $B$. This formulation shows that, within linear elasticity, bending energy can be equivalently expressed as stretching energy between next-nearest neighbors. (For further simulation-based validation, see Ref. (*35*).)

## III. DYNAMIC RESPONSE FOR THE SPRING-MASS MODEL

In this section, we examine the dynamical spectrum of the spring-mass model for the deformed pyrochlore lattice, depicted in Fig. 4(a) of the main text. The lattice comprises $N$ sites, corresponding to $3N$ degrees of freedom. To probe its behavior, we apply a monochromatic excitation at the tip of one tetrahedron on the open surface, oscillating with frequency $\omega$. This external excitation imposes a harmonic displacement $u_c(\omega t)$ on the shaken site, while the remaining sites oscillate freely in response to this boundary-driven motion. We represent their displacements as a $3N \times 1$ vector: $u_f = (u_1, u_2, \ldots, u_{i \neq c}, \ldots, u_N)$. Thus, the total displacement of all sites in the lattice, $u$, can be expressed as: $u = (u_f, u_c)$. The evolution of these displacements follows the Newtonian equation of motion, governing the system's response.

$$m\ddot{\boldsymbol{u}} + \eta\dot{\boldsymbol{u}} + \boldsymbol{D}\boldsymbol{u} = \boldsymbol{0}. \tag{18}$$

Here, $m$ is the mass of the nodes in the theoretical spring-mass model, $\eta$ represents the damping, and $\boldsymbol{D}$ denotes the dynamic matrix as defined in SI. Eq. (9).

We separate the displacement vector into the boundary-driven component (i.e., $u_c$) and the responding component ($u_f$). This leads to the block-matrix formulation of the Newtonian dynamics:

$$\begin{cases} m\ddot{u}_f + \eta\dot{u}_f + D_{ff}u_f + D_{fc}u_c = 0 \\ m\ddot{u}_c + \eta\dot{u}_c + D_{cf}u_f + D_{cc}u_c = 0 \end{cases} \tag{19}$$

where $D_{ff}$ and $D_{cc}$ represent the dynamical matrices for the degrees of freedom in $u_f$ and $u_c$, respectively, whereas $D_{fc}$ and $D_{cf}$ represent the coupling between $u_f$ and $u_c$.

We adopt the harmonic ansatz of the site displacement, namely $u_f = A_f(\omega)e^{\mathrm{i}\omega t}$ and $u_c = A_c(\omega)e^{\mathrm{i}\omega t}$, and substitute it into SI. Eq. (19):

$$\begin{cases} A_f(\omega) = (m\omega^2 - \mathrm{i}\eta\omega - D_{ff})^{-1}(D_{fc}A_c) \\ A_c(\omega) = (m\omega^2 - \mathrm{i}\eta\omega - D_{cc})^{-1}(D_{cf}A_f). \end{cases} . \tag{20}$$

The corresponding Green's function are given by:

$$\begin{cases} \boldsymbol{G}(u_c, u_f; \omega) = (m\omega^2 - \mathrm{i}\eta\omega - D_{ff})^{-1}D_{fc} \\ \boldsymbol{G}(u_f, u_c; \omega) = (m\omega^2 - \mathrm{i}\eta\omega - D_{cc})^{-1}D_{cf}. \end{cases} \tag{21}$$

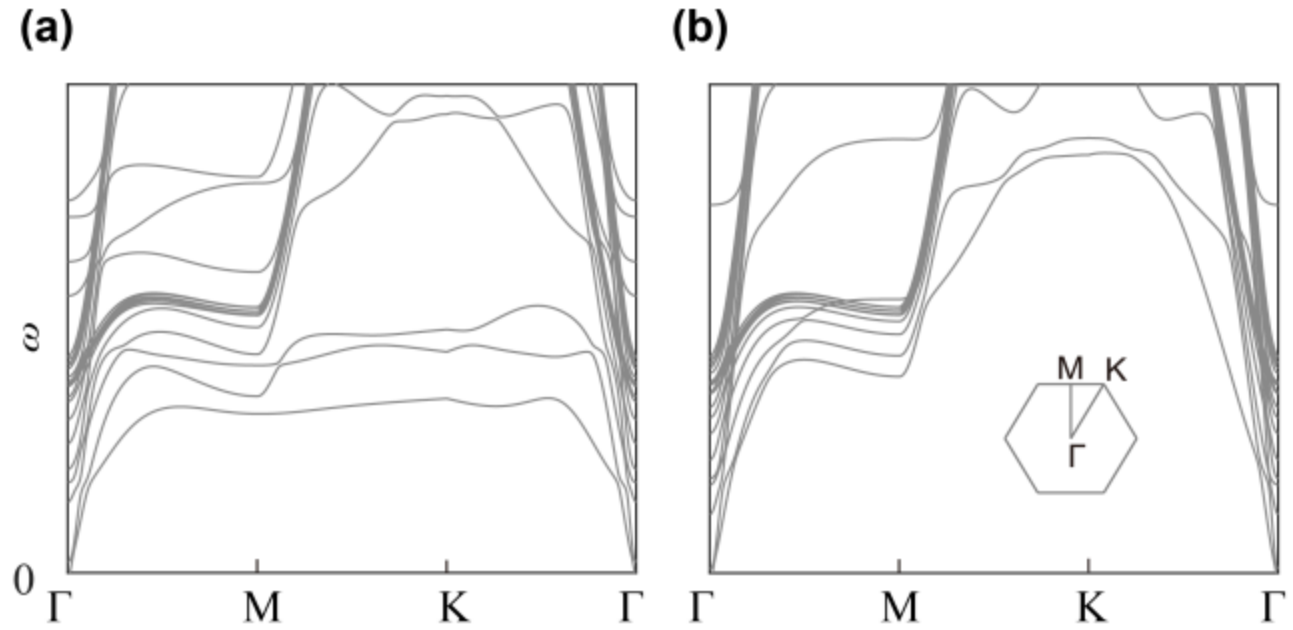

FIG. S3. (a, b) Band structures of the spring-mass model excited at the soft and rigid boundaries, respectively. The parameter is the same as the Fig. 1 in the main text.

Using SI. Eq. (21), we can calculate the dynamic response of the entire system under specific excitations.

We now compute the band structure of the Newtonian dynamics of the spring-mass model, whose supercell is composed of 10 unit cells. Periodic boundary conditions are applied in the x and y directions, and semi-open boundary conditions are applied in the z direction, i.e. when excited at the soft (rigid) boundary, fix the rigid (soft) boundary. The results are illustrated in SI. Fig. S3. As expected, the three branches of topology soft modes arise in the SI. Fig. S3(a) when the excitation signal is imposed on the soft side, whereas for SI. Fig. S3(b), these topological soft modes are absent for the excitation signal is imposed on the rigid side of the sample. The results are in line with Fig. 3(a) and 3(b) in the main text.

Finally, we compute the ratio of kinetic energy between the excitation boundary and the opposing boundary by varying the damping parameter. Specifically, the damping coefficient $\eta_0 = 15$ was selected so that the theoretically computed ratio of kinetic energy reproduces the experimental measurements presented in the main text.

To further examine the impact of damping, we varied the parameter using $\eta = 10^0\eta_0$, $10^1\eta_0$, and $10^2\eta_0$. As shown in SI Fig. S4, two key features emerge: (1) Increasing the damping coefficient decreases the ratio of kinetic energy for both rigid→soft and soft→rigid boundary transmission, since damping reduces the penetration depth of the topological modes. (2) The contrast in the ratio is also diminished at higher damping, because although the topological soft modes remain polarized and localized at the soft edge, stronger damping suppresses this polarization in both directions (soft→rigid and rigid→soft).

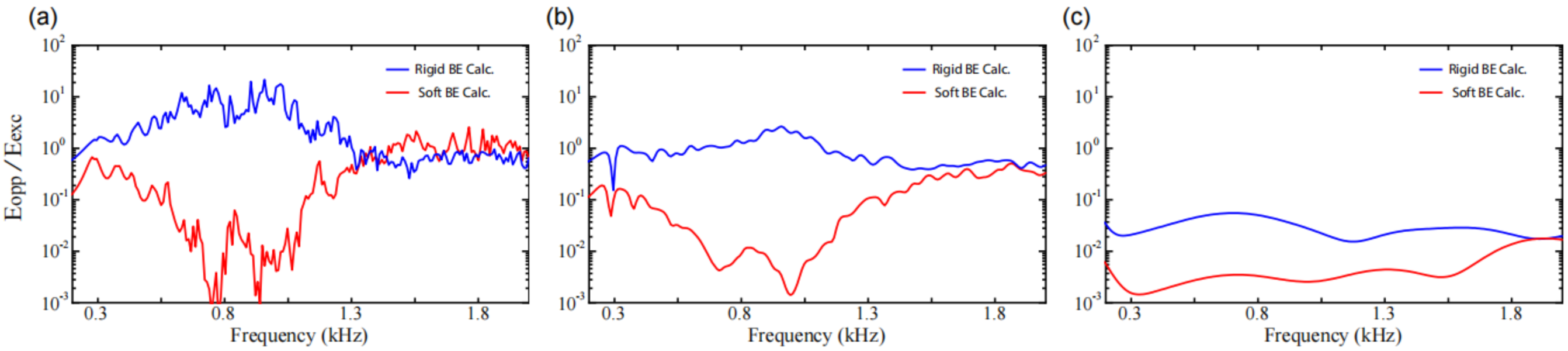

FIG. S4. Numerical results for the ratio of the kinetic energy at the excitation boundary relative to the opposite boundary, shown for damping coefficients $\eta = 10^0\eta_0$, $10^1\eta_0$, and $10^2\eta_0$ in panels (a), (b), and (c), respectively. Here $\eta_0 = 15$ denotes the damping coefficient used in Fig. 4 of the manuscript, selected such that the theoretically computed ratio of kinetic energy reproduces the experimental measurements.

## IV. IMPACT-BASED EXPERIMENT

To more effectively capture the asymmetry in topological boundary elasticity, we introduce an impact-based metric. In the experiment, a rigid metallic sensor probe was released onto the surface, striking either the rigid or the soft boundary of the structure. The resulting force-time response within the structure was recorded using the high-speed force sensor. As shown in SI.Fig. S5, the topologically rigid boundary exhibits a stronger and more concentrated force response to impact, and multiple rebound signals are observed, reflecting its highly localized rigidity. In contrast, the topologically soft boundary produces a markedly smaller force response, indicative of boundary softness that rapidly dissipates the impact forces. This pronounced contrast between the two boundaries further provides a quantitative measure of polarized elasticity.

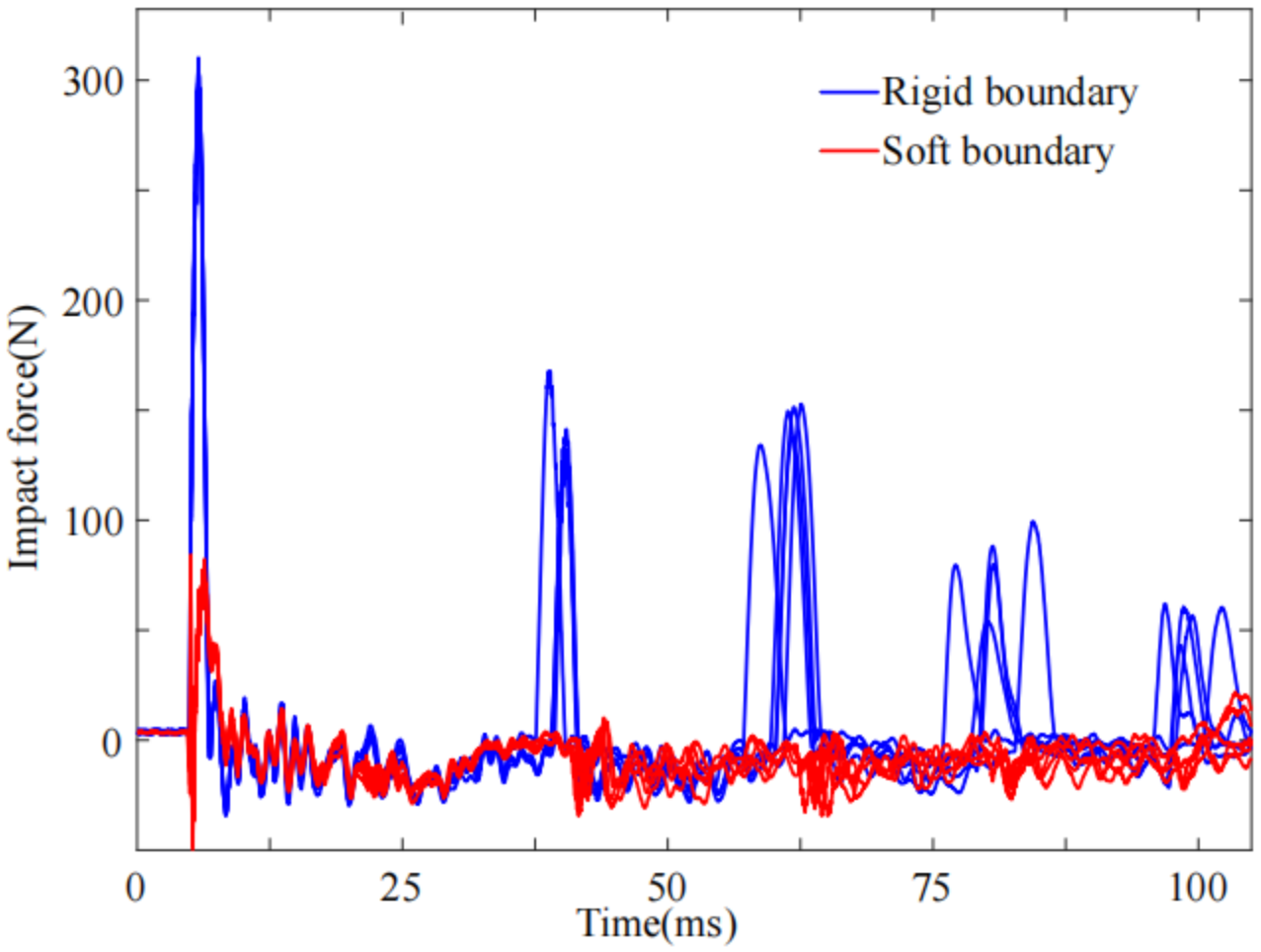

FIG. S5. Impact-based experiment results illustrating the polarized boundary elasticity. Force-time measurements are performed on both the topologically rigid (blue line) and soft (red line) surfaces, with each case repeated 5 times to ensure reproducibility.

## V. SAMPLE FABRICATION

In this section, we introduce the fabrication of the experimental sample.We first discuss the geometric parameters of the unit cell. It consists of two tetrahedra: the vertices of the first tetrahedron are defined based on a deformed pyrochlore lattice, denoted as points $\boldsymbol{A} = (0.5530, 0.5159, 0)l$, $\boldsymbol{B} = (0.0029, 0.5000, 0.5530)l$, $\boldsymbol{C} = (0.4470, 0.1000, 0.4470)l$, and $\boldsymbol{D} = (-0.0954, -0.1000, 0.0636)l$. For the second tetrahedron, point $D$ serves as the shared vertex, while the remaining three vertices are generated by translating points $A$, $B$, and $C$ by the lattice vectors $\boldsymbol{a}_1 = (1, 1, 0)l$, $\boldsymbol{a}_2 = (0, 1, 1)l$, and $\boldsymbol{a}_3 = (1, 0, 1)l$, respectively.And here we take $l = 24\,\text{mm}$ .Under this configuration, the corresponding mass-spring model exhibits a topological polarization vector given by $\boldsymbol{R}_\text{T} = \boldsymbol{a}_1 - \boldsymbol{a}_2 - 2\boldsymbol{a}_3$. As illustrated in Fig. 1 of the main text, the vertices of each tetrahedron are connected by cylindrical elements with a diameter of $D = 4.8\,\text{mm}$, effectively representing the nearest-neighbor Hookean springs. Additionally, to represent next-nearest-neighbor interactions, we introduce flexible hinges at the tetrahedral vertices. These hinges are realized by removing elliptical sections of material around each vertex, possessing a bending stiffness that permits motion within a defined range. As shown in the SI. Fig. S6, the geometry of the elliptical cut is characterized by a semi-major axis $a = 3.6\,\text{mm}$, semi-minor axis $b = 2.16\,\text{mm}$, and a distance $d = 1.8\,\text{mm}$, where $d/2$ represents the distance from the endpoint of the major axis to the vertex of the tetrahedron—corresponding to the radius of the thinnest section of the hinge. The hinge diameter $d$ here plays a crucial role in determining the bending stiffness of the structure: larger values of $d$ result in higher effective spring constants $\kappa$ for the next-nearest-neighbor Hookean interactions. In the limiting case where $d \to 0$, the hinge behaves like an ideal spherical joint, and the system's finite-frequency soft modes morphs back to topological zero modes.

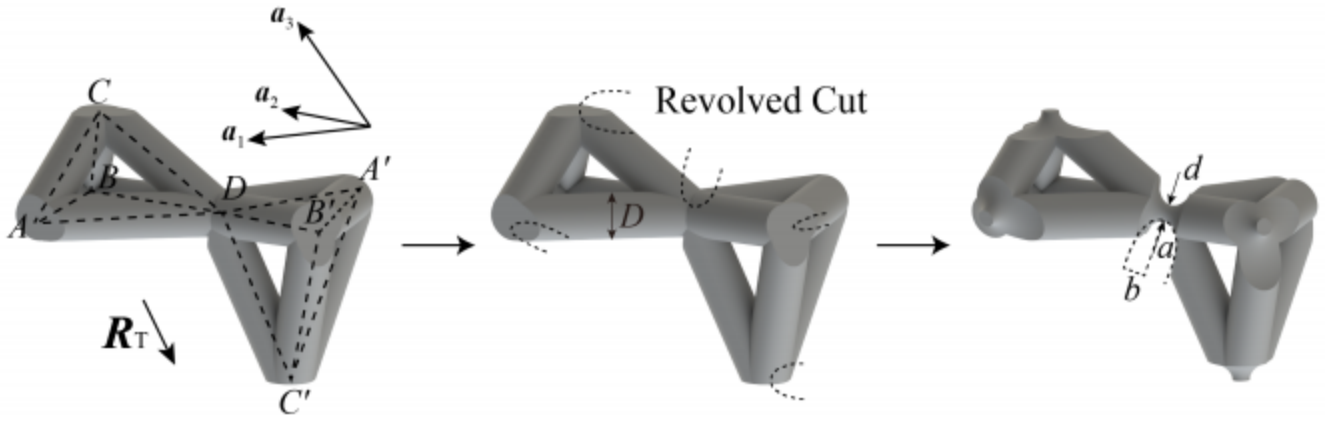

FIG. S6. The details of flexible hinges introduced by removing elliptical sections around each vertex.

We then arrange the unit cells into a periodic lattice. In the experimental realization, a three-dimensional array of $15 \times 15 \times 10$ unit cells is constructed. Open boundary conditions were applied on the top and bottom surfaces, while the four lateral surfaces were enclosed within a rigid boundary shell that was structurally integrated with the lattice. The experimental sample was fabricated using 3D printing technology, employing photosensitive resin molding to achieve the desired geometry and mechanical properties.

## VI. THE STATIC AND DYNAMIC MEASUREMENTS OF THE EXPERIMENTAL SAMPLE

### A. Static measurements

To characterize the sample's response under quasi-static loading, we constructed the experimental setup illustrated in SI. Fig. S7. As depicted, the sample is rigidly fixed to the optical table, while a force sensor is mounted on a vertically movable displacement stage. This sensor makes contact with the central region of the sample. By slowly and continuously actuating the displacement stage in both upward and downward directions, and simultaneously recording data at both the top and bottom boundaries of the sample, we acquire real-time measurements of the response against static force. While maintaining equal loading on the top and bottom boundaries, a laser rangefinder is used to scan a designated region of the sample surface, providing vertical displacement information. This allows us to resolve the differential static responses at the top and bottom boundaries, as presented in Fig. 2(e) and Fig. 2(f) of the main text. Additionally, by synchronously monitoring the vertical displacement of the force sensor probe during the motion of the displacement stage, we extract the force-displacement relationship, as shown in Fig. 2(g).

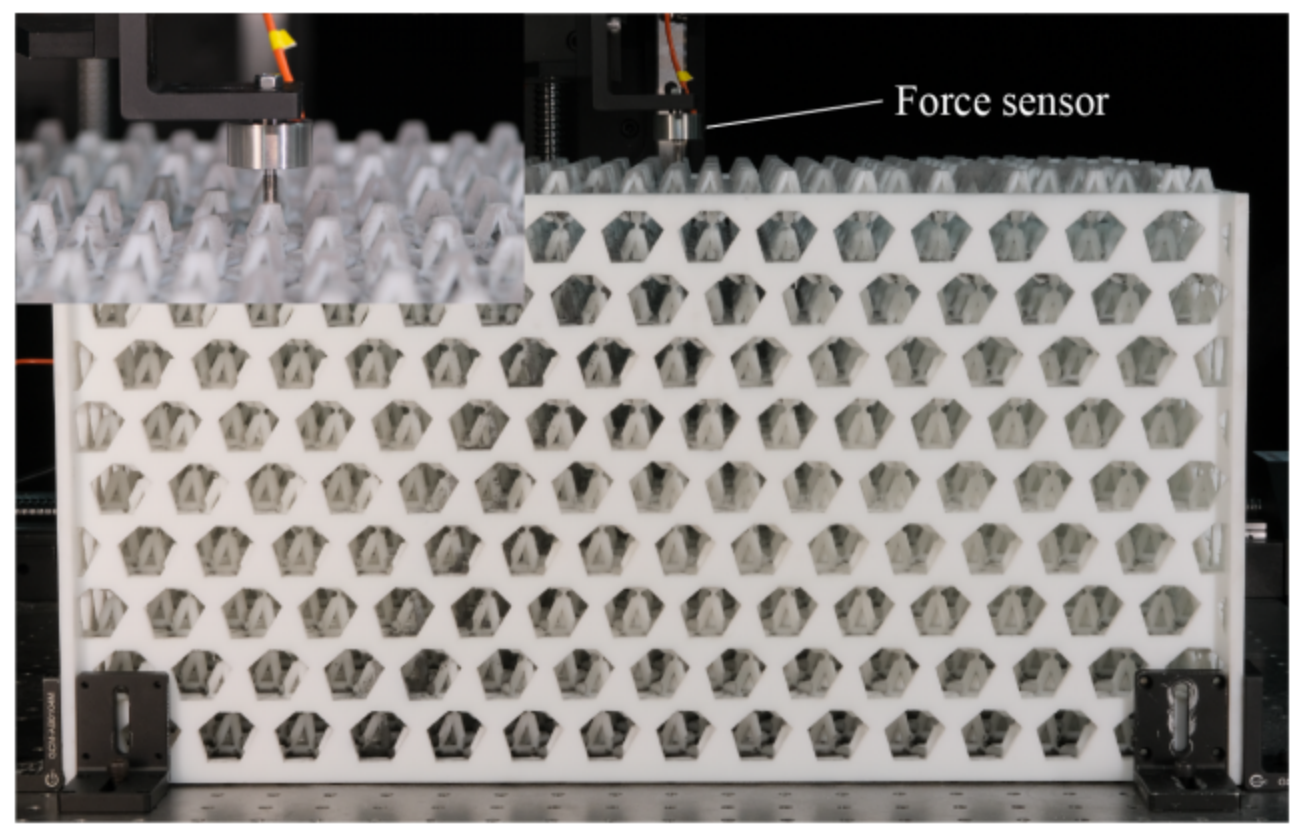

FIG. S7. Experimental setup for static measurements. The force sensor is closely attached to the center of the sample.

### B. Dynamic measurements

To investigate the response of the sample against dynamic external shaking signal, we constructed the experimental setup illustrated in SI. Fig.. S8. As shown, the sample is rigidly fixed to the optical table, and a fixed vibration excitation source is brought into contact with its surface, imposing a displacement boundary condition at a specified frequency. A laser vibrometer, positioned vertically above the sample, is mounted on a displacement platform capable of motion in the horizontal plane.

In the experiment, a chirp signal generated by a network analyzer was used to excite the sample. The frequency range of the excitation spanned from 200 Hz to 2000 Hz, covering the frequency range of the topological edge modes as well as the first two bulk mode dispersions. Simultaneously, the laser vibrometer scanned the entire boundary region of the sample where the excitation was applied. The measured signals were fed back into the network analyzer, enabling extraction of both amplitude and phase information of the frequency-domain response across the scanned frequencies, as presented in Fig. 3(c) and Fig. 3(d) of the main text.

By applying a two-dimensional Fourier transform to the spatially resolved amplitude and phase data in real space, the band structure of the system was reconstructed, as shown in Fig. 3(a) and Fig. 3(b) of the main text. We note that, since the laser vibrometer detects only surface displacements, the experimentally obtained band structure predominantly reflects vibrational modes with significant surface amplitude at both ends of the supercell. The soft-side surface modes have a much greater amplitude compared to the rigid-side surface modes. Topological edge modes are thus clearly observed in the measurements, whereas certain bulk modes at specific frequencies and wave vectors, remain undetected in the experimental band structure.

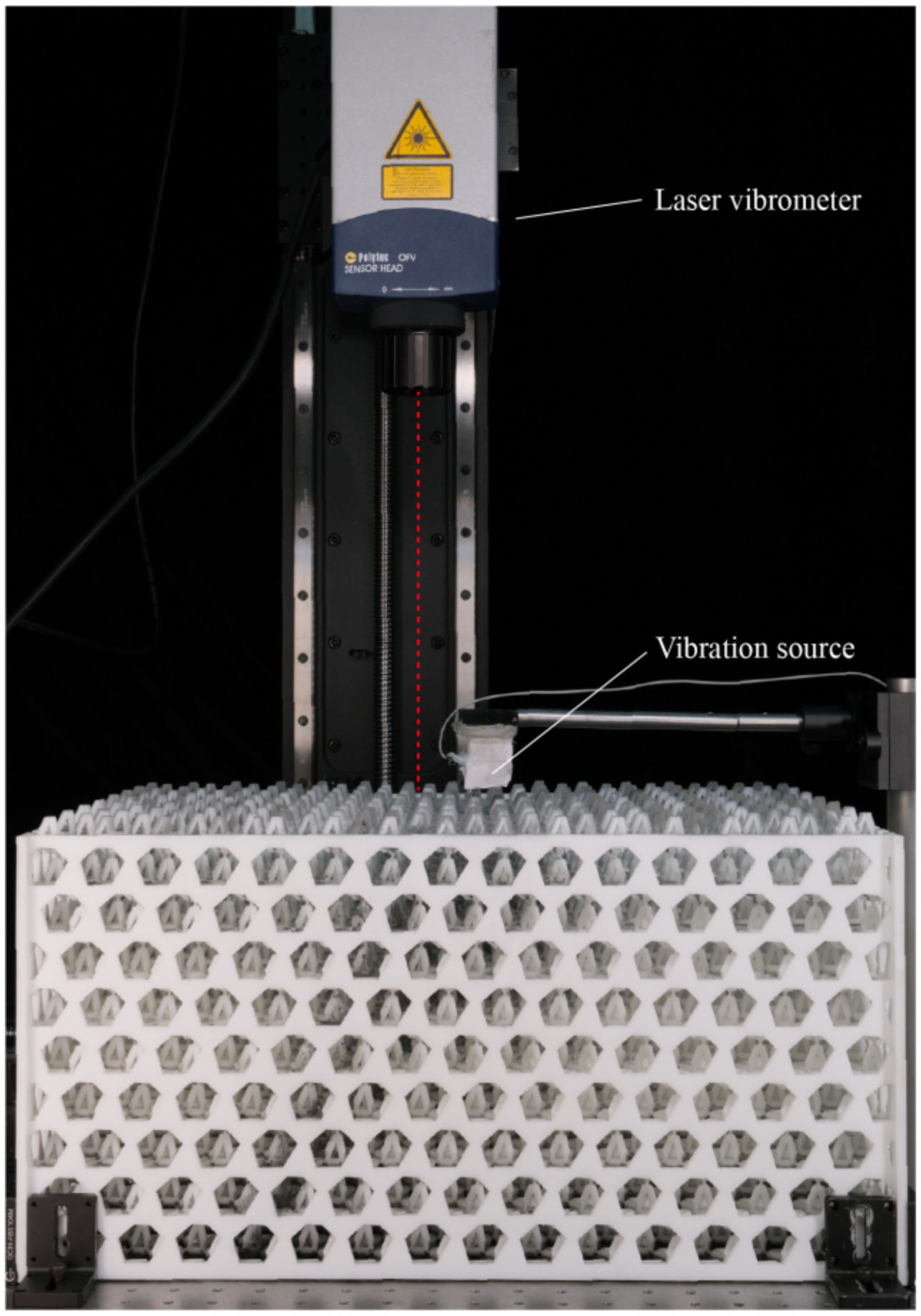

FIG. S8. Experimental setup for dynamic measurements. The vibration source is closely attached to the center of the sample. The red dashed line shows the optical path.

To measure the asymmetric transmission of elastic waves across different frequencies in the sample, we adopt open boundary conditions at both the top and bottom surfaces of the sample. The vibrational excitation source contacts with one side of the sample, while a laser vibrometer is used to scan and record the vibrational response on both the excitation-side boundary and the opposite boundary plane. In this measurement, the amplitude data acquired from both surfaces are squared and summed over all scanned points to quantify the elastic energy on the upper and lower surfaces, respectively. To ensure consistency, the results are normalized by the total number of scanned points on each surface. The ratio between the kinetic energies on the two surfaces is then calculated to determine the frequency-dependent transmissibility, as shown in Fig. 4(a) of the main text.

To characterize the internal response of the sample at specific frequencies, we employed the experimental setup illustrated in SI. Fig. S9, which consists of a coaxial alignment system, a hollow metal tube, and a miniature reflector. The entire assembly is rigidly mounted beneath the laser vibrometer. The device, primarily composed of the hollow metal tube, is inserted into the interior of the sample through its hollowed-out section. By aligning the metal tube coaxially with the laser beam emitted by the vibrometer, the laser passes through the tube and is reflected by the mirror onto the inner surface of the sample. The reflected beam then travels back along the same optical path to the vibrometer, enabling measurement of the internal vibrational response. During the experiment, we performed measurements on all unit cells along the cross-sectional plane where the vibration excitation source was applied. For each unit cell, a measurement point was selected at the same relative position to ensure spatial consistency. This procedure allowed us to reconstruct the spatial profile of the sample's frequency-domain response, as presented in Fig. 4(b) of the main text.

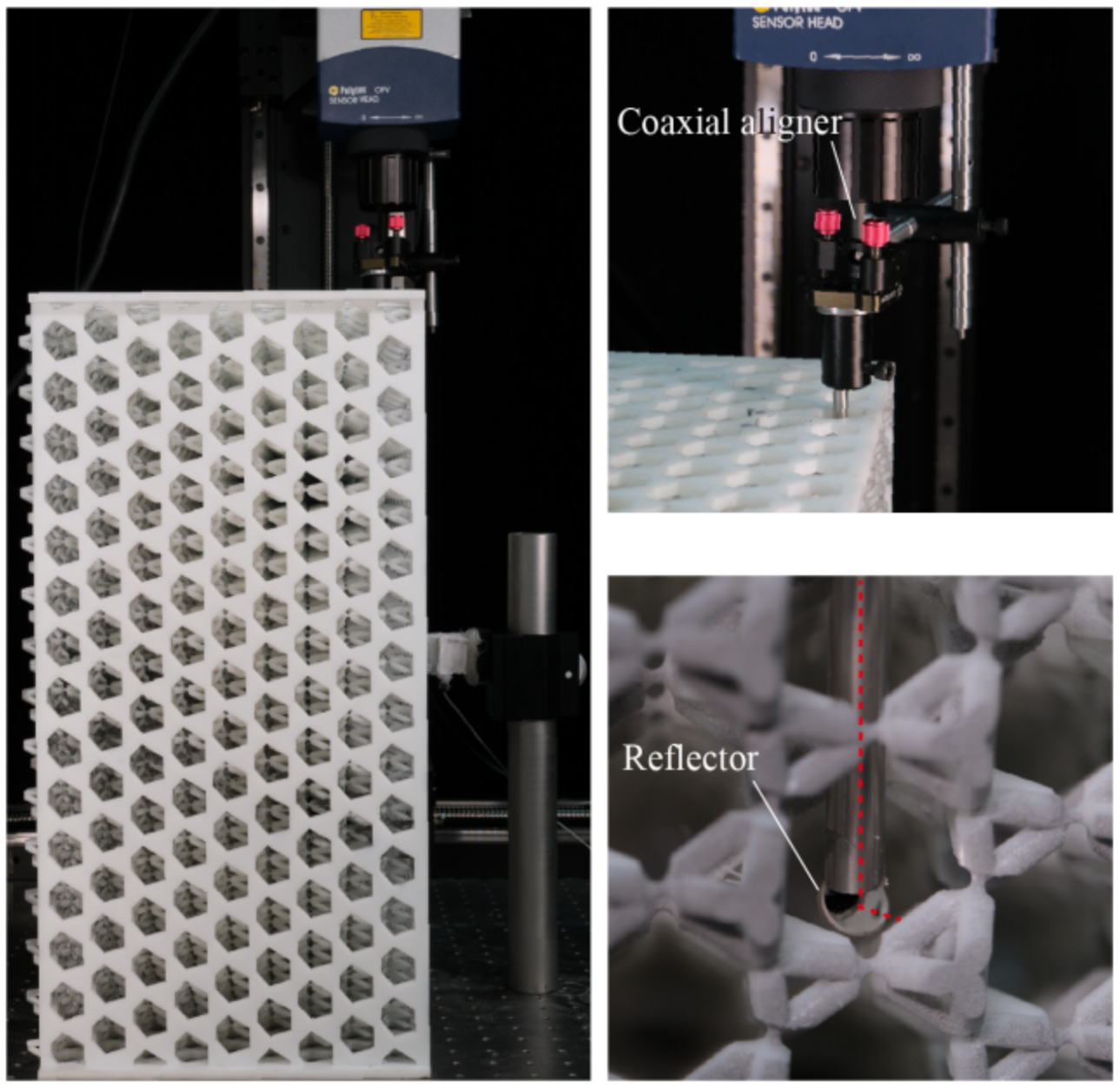

FIG. S9. Experimental setup for the Internal characterization of the sample. The sample is placed vertically. The red dashed line shows the optical path.

### C. Impact-based measurements

To characterize the sample's response to impact, a dedicated test setup was constructed, as illustrated in SI.Fig. S10. A high-speed force sensor (PCB-208C01) was mounted on a linear guide rail, constraining its motion to the vertical axis. An attached ferromagnetic component allowed the entire sensor assembly to be held at a predetermined release height by an electromagnetic switch. The impact was initiated by turning off the electromagnetic switch, which released the assembly to fall freely from a fixed initial position onto the sample. The resulting force signal was synchronously captured by a triggered oscilloscope (KEYSIGHT DSOX4054A), yielding the force-time history of the impact event.

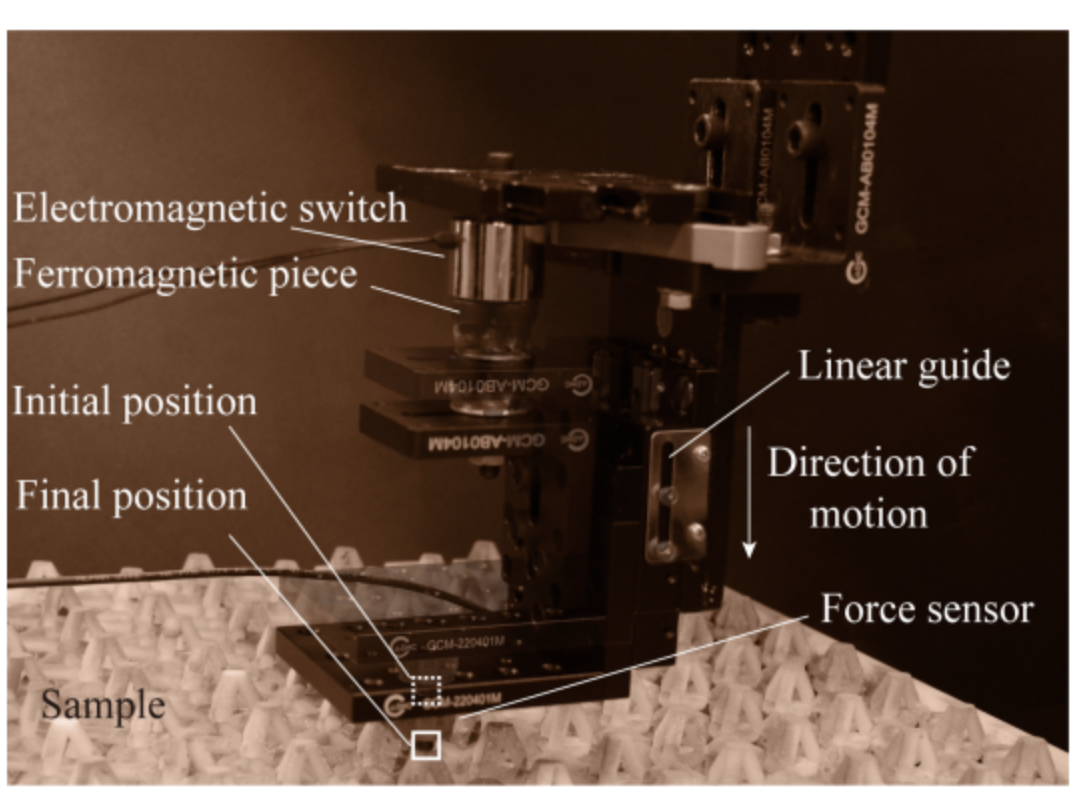

FIG. S10. Experimental set up for impact-based measurements. The mechanical probe impacts the sample in the vertical direction. The dashed and solid boxes indicating the initial and final positions of the motion respectively.